\documentclass{aa}

\usepackage{natbib}
\usepackage{graphicx}
\usepackage{txfonts}
\usepackage{float}
\usepackage{xcolor}
\usepackage[colorlinks=true,citecolor=blue]{hyperref}

\makeatletter
\renewcommand*\aa@pageof{, page \thepage{} of \pageref*{LastPage}}
\makeatother

\begin{document} 
  \title{Downsizing revised: Star formation timescales for elliptical galaxies with an environment-dependent IMF and number of SNIa}
  \titlerunning{SFTs for elliptical galaxies with an environment-dependent IMF and number of SNIa}

  \author{Zhiqiang Yan\inst{1}
          \and
          Tereza Jerabkova\inst{2}\thanks{ESA fellow}
          \and
          Pavel Kroupa \inst{1} \fnmsep \inst{3}
          }

  \institute{
  Charles University in Prague, Faculty of Mathematics and Physics, Astronomical Institute, V Hole{\v s}ovi{\v c}k{\'a}ch 2, CZ-180 00 Praha 8, Czech Republic
  \\ Emails: yan@astro.uni-bonn.de; tereza.jerabkova@esa.int; pkroupa@uni-bonn.de
          \and
  ESTEC/SCI-S, Keplerlaan 1, 2200 AG Noordwijk, Netherlands        
          \and
  Helmholtz-Institut f{\"u}r Strahlen- und Kernphysik (HISKP), Universität Bonn, Nussallee 14–16, 53115 Bonn, Germany
            }

  \date{Received  01.03.2021  / Accepted  07.07.2021}

  \abstract
  {
  Previous studies of the stellar mean metallicity and [Mg/Fe] values of massive elliptical (E)~galaxies suggest that their stars were formed in a very short timescale which cannot be reconciled with estimates from stellar population synthesis (SPS) studies and with hierarchical-assembly. Applying the previously developed chemical evolution code, GalIMF, which allows an environment-dependent stellar initial mass function (IMF) to be applied in the integrated galaxy initial mass function (IGIMF) theory instead of an invariant canonical IMF, the star formation timescales (SFT) of E galaxies are re-evaluated. The code's uniqueness lies in it allowing the galaxy-wide IMF and associated chemical enrichment to evolve as the physical conditions in the galaxy change. The calculated SFTs become consistent with the independent SPS results if the number of type Ia supernovae (SNIa) per unit stellar mass increases for more massive E~galaxies. This is a natural outcome of galaxies with higher star-formation rates producing more massive star clusters, spawning a larger number of SNIa progenitors per star. The calculations show E~galaxies with a stellar mass $\approx 10^{9.5} M_\odot$ to have had the longest mean SFTs of $\approx2\,$Gyr. The bulk of more massive E~galaxies were formed faster (SFT$\,\approx 1\,$Gyr) leading to domination by M~dwarf stars and larger dynamical mass-to-light ratios as observed, while lower-mass galaxies tend to lose their gas supply more easily due to their shallower potential and therefore also have similarly-short mean SFTs. This work achieves, for the first time, consistency of the SFTs for early-type galaxies between chemical-enrichment and SPS modelling and leads to an improved understanding of how the star formation environment may affect the total number of SNIa per unit stellar mass formed.
  }


\maketitle
 


\section{Introduction}\label{sec:intro}

The vast number of galaxies are rotationally-supported disk galaxies with quite constant star-formation histories \citep{Schombert+20, 2020MNRAS.497...37K, Hoffmann+20} and comprise the standard outcome of galaxy formation. Elliptical (E) galaxies are rare pressure supported stellar systems, comprising only a few per cent of the galaxy population \citep{2010A&A...509A..78D}, but they include the most massive galaxies with a dynamical mass (stars and stellar remnants) larger than $M_{\rm dyn} \approx 10^{10}\,M_\odot$. The formation of largely pressure-supported galaxies thus constitutes an interesting problem to solve in a model of cosmological structure formation, particularly since these galaxies are today known to have formed very early after the Big Bang. E~galaxies furthermore have a number of properties that are not yet well understood including their chemical enrichment history.

Dynamical analysis \citep{2021A&A...647A.145P}, spectroscopic studies \citep{2020ApJ...905...40S}, and analysis of resolved galaxies \citep{2021A&A...645L...1B,2021arXiv210407649L} have confirmed that
the central regions of massive early-type galaxies (ETGs) form in a monolithic collapse with a short star-formation time-scale (SFT), and therefore can be reasonably approximated by the `closed-box' model \citep{2012ceg..book.....M} with neither galactic in- nor outflow (we return to this in Section~\ref{sec:Closed-box model}). The central and satellite ETGs show no significant differences in their age and $\alpha$-element gradients, suggesting a similar formation history \citep{2020ApJ...896...75S}.
With a well-constrained galaxy-wide stellar initial mass function (gwIMF), stellar yields, type~Ia supernovae (SNIa) normalization and delay-time-distribution (DTD), the stellar element abundance ratio of a galaxy with its present-day mass is determined in the closed-box model solely by the initial gas supply and the star formation history (SFH). One can then estimate the SFT of a monolithically collapsed E~galaxy through its stellar mean alpha element to iron peak element abundance ratio (e.g. [Mg/Fe]). This was achieved by \citet{2005ApJ...621..673T,2010MNRAS.404.1775T}, who showed that (i) more massive E~galaxies started to form sooner after the Big Bang, (ii) the resulting SFTs constrained by [Mg/Fe], $\tau_{\rm SF, Mg/Fe}(M_{\rm dyn})$, become shorter with increasing $M_{\rm dyn}$ (comprising the downsizing problem as it is contrary to the expectation that more massive galaxies need longer to form through mergers in the standard dark-matter based cosmological models), and (iii) the $\tau_{\rm SF, Mg/Fe}(M_{\rm dyn})$ values are shorter than the $\tau_{\rm SF, SPS}(M_{\rm dyn})$ values obtained by stellar population synthesis (SPS) studies by, e.g., \citet{2015MNRAS.448.3484M} and \citet{2020A&A...644A.117L}.
On the other hand, dark-matter cosmology-motivated hydrodynamical simulations of the hierarchical formation of the most massive ellipticals lead to the result that the empirical $\tau_{\rm SF, Mg/Fe}$ values are too short (less than a Gyr); there not being enough time for the synthesised and released elements to recycle and increase the mean stellar metallicity to the observed level \citep{2009A&A...499..409C,2009A&A...505.1075P,2017MNRAS.466L..88D,2017MNRAS.464.4866O,2020ApJ...897L..42J}.

The situation worsens when one considers jointly the galactic stellar metallicity in addition to [Mg/Fe] in a chemical evolution model due to the metal-rich stars having a lower [Mg/Fe] yield (see e.g. \citealt[their section 2.1.3]{2012ceg..book.....M}).
This has been studied in \citet{2004MNRAS.347..968P} where three different masses of galaxies are modelled (shown in Fig.~\ref{fig:best_fit_SFT} below) 
although the fits of observed E~galaxies are not ideal (cf. \citealt{2011A&A...530A..98P}).
With the standard assumption of an invariant and canonical gwIMF \citep{1955ApJ...121..161S,2001MNRAS.322..231K} and using closed-box modelling, \citet{2019A&A...632A.110Y} obtained a good fit to the metallicity and alpha element abundances but the implied $\tau_{\rm SF}$--$M_{\rm dyn}$ relation is steeper than the relation suggested originally by \citet{2005ApJ...621..673T} regardless of a potential systematic bias on the observed metal abundances or on the stellar magnesium yield uncertainty. This revised $\tau_{\rm SF}$--$M_{\rm dyn}$ relation suggests an even shorter $\tau_{\rm SF}$ for massive galaxies than that implied by the SPS studies and poses an even more severe downsizing problem unable to resolve in the standard cosmological hydrodynamical simulations.
A different gwIMF and/or different stellar yields, different SNIa delay-time distributions (DTDs), gas mixing and expulsion physics, or a combination of these are probably needed to solve this problem.

The most promising solution is to apply a systematically varying gwIMF. With an increasing number of observational studies supporting this idea, it has become a crucial aspect in a galaxy chemical evolution model and needs to be investigated, given the above problems between theory and observation. Other possible solutions assuming an invariant canonical stellar initial mass function (IMF) are discussed in Section~\ref{sec: alternative solutions}.
The gwIMF appears to be top-heavy (containing more massive stars) when the star formation rate (SFR) is high  \citep{2011MNRAS.415.1647G,2018Natur.558..260Z} and top-light for low-SFR dwarf galaxies \citep{2018A&A...620A..39J}. The gwIMF of low-mass stars is also variable. It becomes bottom-light (containing fewer low-mass stars) in metal-poor environments \citep{2018ApJ...855...20G,2020A&A...637A..68Y} and bottom-heavy for the super-solar-metallicity E~galaxies (e.g. \citealt{2021MNRAS.500.3368S}). This means the gwIMF would change over time depending on the SFH and metal enrichment history, which helps to build up the high metallicity and [Mg/Fe] ratio for the massive E~galaxies. The need for a time-variable gwIMF to explain the galactic abundance has been argued in, for example,
\citet{1997ApJS..111..203V}, \citet{1998MNRAS.301..569L}, \citet{2013MNRAS.435.2274W}, \citet{2013MNRAS.436.2892N}, \citet{2013ApJ...779....9B}, \citet{2015MNRAS.448L..82F}, and \citet{2018MNRAS.475.3700M}. See also the reviews given by \citet{2018PASA...35...39H} and \citet{2020ARA&A..58..577S} on other reasons for considering IMF variations.

The applications of environment-dependent gwIMFs do show promising results \citep{2009A&A...499..711R,2017MNRAS.464.3812F,2018MNRAS.479.5448B,2019MNRAS.483.2217D,2019MNRAS.482..118G,2020MNRAS.494.2355P,2020A&A...637A..68Y}. 
But it is important to use formulations of the gwIMF which are consistent with the extragalactic data and, at the same time, are also consistent with observed resolved stellar populations for both massive and low-mass stars (detailed in Section~\ref{sec:IGIMF}). With this in mind, for the present work, we focus on the advanced variable gwIMF theory with various verified predictions, i.e., we apply the integrated galaxy-wide IMF (IGIMF) theory 
\citep{2003ApJ...598.1076K,2011MNRAS.412..979W,2017A&A...607A.126Y,2018A&A...620A..39J,2020A&A...637A..68Y} to compute the gwIMF. The IGIMF theory is a framework that accounts for the gwIMF being made up of the IMFs of all embedded clusters forming in the galaxy \citep{2003ApJ...598.1076K} and applies the empirical rules which modify the stellar IMF in embedded clusters according to the metallicity and density of an embedded cluster \citep{2012MNRAS.422.2246M, 2018A&A...620A..39J}, being supported by the recent study of \citet{2017ApJ...850L..14V} and \citet{2019A&A...626A.124M}.

Thus, here we apply, for the first time, the IGIMF theory to the monolithic E~galaxy chemical evolution model, study how it affects the SFTs of E~galaxies, whether the result is consistent with SPS studies, and what the implications might be. 
The open-source chemical evolution model developed by \citet{2019A&A...629A..93Y} is applied, accounting specifically for the strong gwIMF variation and calculating the element enrichment self-consistently with the number of SNIa affected by the IMF. 
The long-term aim of this research project is to ultimately understand how hydrodynamical models of galaxy formation might be able to be made consistent with the observational constraints (mass, alpha-element abundances, and metallicities of gas and stars) of E~galaxies. The closed-box models studied here are to be seen as a comprehensive parameter study, which will allow future computer-costly fully-self consistent galaxy formation computations to be performed using this knowledge gain.

Section~\ref{sec:Model} summarises the method used here to calculate the gwIMF and the chemical evolution of a galaxy. In particular, we emphasise how the number of SNIa is affected by the IMF shape.
The observed chemical abundances of elliptical galaxies are briefly introduced in Section~\ref{sec:Observational constrain}.
Then the likelihood of certain SFTs for a galaxy can be determined by comparing the observed and modelled element abundances of galaxies as is detailed in Section~\ref{sec:Method}.
The resulting most-likely $\tau_{\rm SF}$--$M_{\rm dyn}$ relations adopting different IMF and SNIa formulations are shown in Section~\ref{sec:Results}. 
The reliability of the assumptions applied in our model is discussed in Section~\ref{sec:Discussions}. 
Finally, Section~\ref{sec:Conclusion} contains our conclusions.

\section{Galaxy model}\label{sec:Model}

This section introduces the GalIMF code which combines the IGIMF theory (Section~\ref{sec:IGIMF}) with the galaxy chemical evolution model (Section~\ref{sec:chemical model}). Following \citet{2019A&A...629A..93Y}, we introduce the mathematical formulation that calculates the number of SNIa in Section~\ref{sec:SNIa} and demonstrate how the number of SNIa is affected by the IMF variation.

\subsection{The systematically varying gwIMF}\label{sec:IGIMF}

The IGIMF theory describes how the gwIMF should vary as a function of galactic properties and is computed by assuming that in each star-formation epoch of duration $\delta t$ in which the average SFR\footnote{Throughout this manuscript, the SFR is in units of $M_\odot/$yr.} is $\bar{\psi}_{\delta t}$, the galaxy forms a total mass in stars, $M_{\rm tot}=\bar{\psi}_{\delta t}\,\delta t$, and that this mass is distributed over a fully populated mass function of embedded star clusters, the ECMF.\footnote{"Fully populated" means the ECMF, in this case, is optimally sampled \citep{2013pss5.book..115K, 2015A&A...582A..93S}.} The time-scale $\delta t = 10\,$Myr, discussed in \cite{2015A&A...582A..93S}, is essentially the lifetime of molecular clouds and is also the time-scale in which the inter-stellar-medium of a galaxy churns out a full population of freshly formed embedded clusters, each of which dissolves into the galactic field through gas expulsion, stellar evolution mass loss and two-body relaxation-driven evaporation.

The IGIMF theory is based on a set of axioms that are formulated based on observational constraints (e.g. \citealt{2015MNRAS.446.4168R}). These include how the stellar IMF and the ECMF change 
\citep{2003ApJ...598.1076K,2011MNRAS.412..979W,2018A&A...620A..39J,2020A&A...637A..68Y}. 
The IGIMF theory has solved a number of previously outstanding extragalactic problems such as explaining the UV extended galactic disks \citep{2008Natur.455..641P}, predicting the diverging H$\alpha$- vs UV-fluxes of dwarf galaxies (\citealt{2009MNRAS.395..394P}), verified by \citet{2009ApJ...706..599L}, a lower $\alpha$ element to iron peak element ratio in dwarf galaxies \citep{2020A&A...637A..68Y,2020A&A...642A.176T,2021ApJ...910..114M}, and naturally accounting for the time-scale problem for building up a sufficient stellar population in dwarf galaxies given their low SFRs \citep{2009ApJ...706..516P}. In the following, we specify the exact formula for calculating the gwIMF that is applied for this work.

The IMF for stars with a mass higher than $1 M_\odot$ (Eq.~\ref{eq:alpha3} below) follows the prescription given in \citet{2017A&A...607A.126Y}, while the IMF of low-mass stars (i.e. $\alpha_1$ and $\alpha_2$ in Eq.~\ref{eq:IMF}) follows equation~9 of \citet{2020A&A...637A..68Y}. We have for the stellar IMF ($m$ is in unit of solar mass)
\begin{equation}\label{eq:IMF}
    \xi_{\star} (m) =\mathrm{d} N_{\star}/\mathrm{d} m=    \left\{ \begin{array}{ll}
    2k_{\mathrm{\star}} m^{-\alpha_1}, \hspace{0.5cm} 0.08\leq m/M_{\odot}<0.50 \,, \\
    k_{\mathrm{\star}} m^{-\alpha_2}, \hspace{0.5cm} 0.50\leq m/M_{\odot}<1.00 \,, \\
    k_{\mathrm{\star}} m^{-\alpha_3}, \hspace{0.5cm} 1.00\leq m/M_{\odot}< m_{\mathrm{max}} \,, \\
    \end{array} \right.
\end{equation}
where the number of stars in the mass interval $m$ to $m+dm$ is $dN_{\star}$. We note that the normalization parameter for stars in the smallest mass interval, $2k_{\mathrm{\star}}$, is two times the normalization parameter for more massive stars because we presume that IMF is a continuous function and that $\alpha_2-\alpha_1=1$ always holds (see Eq.~\ref{eq:alpha18} below). The embedded-cluster-mass-dependent stellar mass upper limit, $m_{\rm max}(M_{\rm ecl}) \le m_{\rm max*}=150\,M_\odot$, and the normalization parameter, $k_\star$, is determined by simultaneously solving for the mass in stars formed in the embedded cluster,
\begin{equation}
M_{\mathrm{ecl}}=\int_{0.08~\mathrm{M}_{\odot}}^{m_{\mathrm{max}}}m~\xi_{\mathrm{\star}}(m)\,\mathrm{d}m,
\end{equation}
and the statement that there is one most massive star in this embedded cluster,
\begin{equation}\label{eq:normal IMF}
    1=\int_{m_{\mathrm{max}}}^{m_{\rm max*}} \xi_{\star}(m)  \,\mathrm{d}m,
\end{equation}
$m_{\rm max*}$ being the adopted physical upper mass limit, $150 M_\odot$ \citep{2017A&A...607A.126Y}.

The power-law indices or slopes of the IMF for low- and intermediate-mass stars is determined empirically by \citet{2020A&A...637A..68Y} as
\begin{equation}\label{eq:alpha18}
\begin{split}
    &\alpha_1=1.3+\Delta\alpha \cdot (Z-Z_\odot),\\
    &\alpha_2=2.3+\Delta\alpha \cdot (Z-Z_\odot),
\end{split}
\end{equation}
where, $Z$ and $Z_\odot=0.0142$ are, respectively, the mean stellar metal mass fraction for the target system and the Sun. The value of $\Delta\alpha=35$ suggested by \citet{2020A&A...637A..68Y} in their equation 9 is now updated to $\Delta\alpha=63$ due to a different $Z_\odot$ assumption (\citealt{2020A&A...637A..68Y} should also apply 0.0142 but actually applied 0.02 in the code; this change negligibly affecting the results). We note that this latest IGIMF formulation is different from the one applied in \citet{2017MNRAS.464.3812F} where the IMF for low- and intermediate-mass stars are invariant. Evidence for a systematic change of $\alpha_1$ and $\alpha_2$ with metallicity was already noted by \citet{2001MNRAS.322..231K} and formulated in \citet{2002Sci...295...82K}, \citet{2012MNRAS.422.2246M}, and \citet{2014MNRAS.437..994R}. Recent observations of massive ETGs also support a bottom-heavy IMF for galaxies with super-solar metallicity \citep{2019MNRAS.485.5256Z,2020ARA&A..58..577S}, in agreement with Eq.~\ref{eq:alpha18}. This is an important factor determining the variable `overall SNIa realisation parameter' introduced in Section~\ref{sec:SFT with variable SNIa rate} below.

The variation of the power-law indices of the IMF for massive stars is determined empirically by \citet{2012MNRAS.422.2246M} as
\begin{equation}\label{eq:alpha3}
\alpha_3=
        \begin{cases} 
            2.3, & x<-0.87, \\
            -0.41x+1.94, & x>-0.87.
        \end{cases}
\end{equation}
The parameter $x$ in Eq.~\ref{eq:alpha3} depends on the metallicity,
$[Z/X]=\mathrm{log}_{10}(Z/X)-\mathrm{log}_{10}(Z_\odot/X_\odot)$, where $Z$ and $X$ are the initial metal and hydrogen mass fraction in stars, respectively. We assume the initial hydrogen mass fraction of any star, $X$, is the same as the Sun, i.e., $X\approx X_\odot$, therefore, $[Z/X]\approx \mathrm{log}_{10}(Z/Z_\odot)=[Z]$. In addition, the parameter $x$ in Eq.~\ref{eq:alpha3} depends on the core density (final stars plus residual gas assuming a star-formation efficiency of $1/3$), $\rho_{\mathrm{cl}}$, of the molecular cloud core which forms the embedded star cluster,
\begin{equation}\label{eq:xxx}
    x=-0.14[\mathrm{Z}]+0.99\log_{10}(\rho_{\mathrm{cl}}/10^6)
\end{equation}
with
\begin{equation}
    \log_{10}\rho_{\mathrm{cl}}=0.61\log_{10}M_{\rm ecl}+2.85,
\end{equation}
as is explained and applied in \citet[their eq. 7]{2017A&A...607A.126Y}, with $\rho_{\rm cl}$ and $M_{\rm ecl}$ being in astronomical units. For example, an embedded star cluster weighing $M_{\rm ecl}=10^3\,M_\odot$ in stars has $\rho_{\rm cl}=4.79\times 10^4\,M_\odot/$pc$^3$.

The distribution of all embedded star clusters formed in a galaxy within the $\delta t=10\,$Myr star-formation epoch is approximated by a power-law ECMF \citep{2006A&A...450..129G,2017MNRAS.465.3775L},
\begin{equation}\label{eq:xi_ecl}
\xi_{\mathrm{ecl}}=\mathrm{d} N_{\rm ecl}/\mathrm{d} M_{\rm ecl}=
k_{\mathrm{ecl}} M_{\rm ecl}^{-\beta}, \hspace{0.5cm} 5 M_\odot \leqslant M_{\rm ecl} <M_{\mathrm{ecl,max}}.
\end{equation}
The total mass formed in stars during the $\delta t$ epoch is given by
\begin{equation}
M_{\mathrm{tot}}=\int_{M_{\mathrm{ecl,min}}}^{M_{\mathrm{ecl,max}}}M_{\rm ecl} \, \xi_{\mathrm{ecl}}(M_{\rm ecl})\,\mathrm{d}M_{\rm ecl} =\bar{\psi}_{\delta t}\, \delta t,
\end{equation}
and there is one most massive embedded cluster,
\begin{equation}\label{eq:normal ECMF}
1=\int_{M_{\mathrm{ecl, max}}}^{10^9~\mathrm{M}_{\odot}}k_{\mathrm{ecl}} M_{\rm ecl}^{-\beta}\,\mathrm{d}M_{\rm ecl},
\end{equation}
with $10^9 M_\odot$ being about the mass of the most-massive ultra-compact-dwarf galaxy. Solving the above two equations yields $k_{\rm ecl}$ and $M_{\rm ecl,max}$. Note that $M_{\rm ecl,max}=M_{\rm ecl,max}(\bar{\psi}_{\delta t})$ which is a relation that is consistent with the extragalactic most-massive-very-young star clusters vs galaxy-wide SFR data \citep{2004MNRAS.350.1503W, 2015A&A...582A..93S, 2013ApJ...775L..38R}.
The slope of the ECMF, $\beta$, depends on the mean galaxy-wide SFR over the $\delta t$ time period, $\bar{\psi}_{\delta t}$ \citep{2013MNRAS.436.3309W},
\begin{equation}\label{eq:beta-SFR}
\beta=-0.106\log_{10} \bar{\psi}_{\delta t} +2.
\end{equation}

The environment-dependent gwIMF (i.e., the IGIMF) follows by adding up all the stars formed in the $\delta t$ epoch,
\begin{equation}\label{eq:IGIMF}
\xi(t) = \mathrm{d} N_{\star}/\mathrm{d} m=\int_{5 M_\odot}^{M_{\rm ecl,max}} \xi_{\mathrm{\star}}(m,M_{\rm ecl},[Z/X]) \, \xi_{\mathrm{ecl}}(M_{\rm ecl},\bar{\psi}_{\delta t})\,\mathrm{d}M_{\rm ecl}.
\end{equation}
The environment-dependence comes in through the SFR, which is a function of time, $\bar{\psi}_{\delta t} = \bar{\psi}_{\delta t}(t)$, and through the stellar IMF and the ECMF being functions of the time-changing metallicity and SFR.

In summary, Eq.~\ref{eq:IMF}, \ref{eq:alpha18}, \ref{eq:alpha3}, \ref{eq:xxx}, \ref{eq:xi_ecl}, and \ref{eq:beta-SFR} are empirical formulations. 
The integral of Eq.~\ref{eq:normal IMF} and \ref{eq:normal ECMF} being 1 is an ansatz of the optimal sampling theory (\citealt[their section 2.2]{2013pss5.book..115K} and \citealt{2015A&A...582A..93S}).
The existence of the $m_{\rm max} = m_{\rm max}(M_{\rm ecl})$ relation is supported by  observed $m_{\rm max}$--$M_{\rm ecl}$ data as is demonstrated in \citet{2017A&A...607A.126Y} and \citet{2018MNRAS.481..153O}.
The above formulation is thus consistent with the observed constraints on the IMF in resolved star clusters and the Galactic field as well as with the above mentioned extragalactic data and thus fulfils this necessary condition.

\subsection{Number of type Ia supernovae}\label{sec:SNIa}

The number of SNIa depends on the number of possible SNIa progenitors, i.e. on the number of stars with a mass between about 3 to 8 $M_\odot$ (see below). Therefore, the number of SNIa per unit mass of stars formed is a function of the IMF or the gwIMF. In previous research, the SNIa incidence has been taken to correlate with the total number of stars. Such an analytical formulation, first developed by \citet{2005A&A...441.1055G}, assumes the canonical universal IMF, and cannot be directly applied when assuming a different or evolving gwIMF.

Here we adjust the \citet{2005A&A...441.1055G} formulation in order to account for the variable gwIMF. The total number of SNIa explosions for a simple stellar population (SSP, stars formed at the same time) after $t$ years of the birth of the SSP per unit stellar mass of the SSP (i.e. 
the time-integrated number of SNIa per stellar mass formed until time $t$) is 
\begin{equation}\label{eq:SNIa rate}
    n_{\rm Ia}(t, \xi, \bar{\psi}_{\delta t})=N_{\rm Ia}(\xi, \bar{\psi}_{\delta t}) \int_{0}^tf_{\rm delay}(t)\mathrm{d}t \; ,
\end{equation}
where $N_{\rm Ia}$ is the total number of SNIa per unit mass of stars formed in the SSP. The integral from $t=0$ to $\infty$ is 1, therefore, $N_{\rm Ia}=n_{\rm Ia}(t=\infty)\approx n_{\rm Ia}(t=10~\mathrm{Gyr})$. It depends on the gwIMF of the stellar population, $\xi$, and on the star formation environment represented by the SFR, $\bar{\psi}_{\delta t}$. Here $f_{\rm delay}$ is the DTD of the SNIa events for which we adopt the empirical power-law function given by \citet{2012PASA...29..447M},
\begin{equation}\label{eq:DTD}
    f_{\rm delay}(t) =
    \begin{cases} 
        0, & t \leqslant 40~{\rm Myr}, \\
        k \cdot t^{-1}, & t>40~{\rm Myr},
    \end{cases}
\end{equation}
where $f_{\rm delay}(t)$ is the fraction of exploded SNIa for a SSP with age $t$. The normalization factor $k$ is determined by the condition $\int_{t=0}^{\infty}f_{\rm delay}(t)\mathrm{d}t=1$.

In the present context, in which a systematically varying IMF/gwIMF is applied, the total number of SNIa per unit stellar mass of the SSP, $N_{\rm Ia}$, (a.k.a. SNIa production efficiency) is calculated as
\begin{equation}\label{eq:N_Ia}
    N_{\rm Ia}(\xi, \bar{\psi}_{\delta t}) = \frac{n_{3,8}(\xi)}{M_{0.08,150}(\xi)} \cdot B_{\rm bin}(\bar{\psi}_{\delta t}) \cdot \frac{n_{3,8}(\xi)}{n_{0.08,150}(\xi)} \cdot C_{\rm Ia}(\bar{\psi}_{\delta t}),
\end{equation}
where $n$ is the number of stars within the mass range given by its subscript (in the unit of $M_\odot$). Similarly, $M$ is the mass of stars in a given mass range indicated by its subscript. Hence, the first term represents the number of stars that are possible to become SNIa progenitors (the primary star in a binary system) per unit stellar mass of the SSP. The fraction of stars in the initial mass range 3 to 8~$M_\odot$ that eventually explode as SNe Ia consists of the following three terms.
The second term, $B_{\rm bin}$, denotes the binary fraction of the SSP which can be assumed to be a constant or a function of the environment represented by the galaxy-wide SFR, $\bar{\psi}_{\delta t}$. The third term then represents the likelihood of a binary system to have the companion star also in the suitable stellar mass range thus being a potential SNIa progenitor system (a binary white dwarf, WD). 
Both, the first and third terms depend on the IMF of each star cluster since it is extremely unlikely for the binaries to form outside a star cluster. However, since the gwIMF is the mass-weighted sum of all IMFs of individual star clusters in a galaxy, we apply gwIMF on Eq.~\ref{eq:N_Ia} to simplify the calculation when considering a galaxy.
Finally, $C_{\rm Ia}$ is the realisation probability of a SNIa explosion for the potential SNIa progenitor system, that is, the fraction of the above selected binary systems which are able to give rise to a SNIa explosion. Again, it has been assumed to be a constant in previous chemical evolution studies but should, in principle, be a function of the star formation environment because it depends on the distribution of semi-major axes and of the mass ratios of the binary WD systems. 
When the galactic SFR is high, more massive clusters form as calculated by the IGIMF theory (Eq.~\ref{eq:xi_ecl} to \ref{eq:beta-SFR}). Massive clusters lead to substantial long-term dynamical processing of the binary-star population \citep{1975MNRAS.173..729H,2011MNRAS.417.1684M}. The soft initial binaries, which have binary-orbital velocities slower than the velocity dispersion in the cluster, are disrupted, while hard binaries (with binary-orbital velocities that are faster than the velocity dispersion in the cluster) are not disrupted but sink to the cluster centre through dynamical friction. This is also true for WD--stellar or WD--WD binaries. There they harden through encounters which also increase their binary-orbital eccentricities. Few would be expected to be ejected in dynamical strong encounters and these are likely to merge and explode as SNIa. Many will not be able to escape from their massive clusters and will continue encountering other stars and remnants to also ultimately explode as SNIa \citep{2002ApJ...571..830S}. Massive star clusters (and thus high SFRs) are thus expected to correlate with enhanced SNIa events. This appears to be born out, not only by the here-derived Eq.~\ref{eq:N_Ia} but also by \citet{2018MNRAS.479.3563F,2021MNRAS.502.5882F} finding evidence for a significantly enhanced occurrence of SNIa in galaxy clusters.

Regarding the IMF dependent terms in Eq.~\ref{eq:N_Ia}, we assume that the stars with a mass between 3 to 8 $M_\odot$ have an equal probability of leading to a SNIa event, while stars with other masses have zero probability. Stars of, for example, 1.5 $M_\odot$ may also become WDs in reality but, compared to the more massive stars, they have a significantly longer lifetime and almost no chance to be primary stars in binary systems with a total mass above 3 $M_\odot$, which is considered as the lower binary system mass limit for SNIa events \citep{2001ApJ...558..351M}. As a simplification, we apply the same mass limit (3 to 8 $M_\odot$) for the primary and the secondary star. These assumed mass limits affect the dependency of the SNIa number on the IMF insignificantly. Changing both SNIa progenitor mass limits for the primary and the secondary star to, e.g., 2 to 8 $M_\odot$ would slightly increase the SNIa production efficiency for the low-SFR galaxies (shown by Fig.~\ref{fig:SNIa_variation_IGIMF_2t8} in the Appendix) but would not affect our conclusions. Having different mass limits for the primary and the secondary stars would give similar results (Fig.~\ref{fig:SNIa_variation_IGIMF_3a15t8}). 

Therefore, for the present study, The only free parameter in Eq.~\ref{eq:SNIa rate} to Eq.~\ref{eq:N_Ia} is the product $B_{\rm bin}(\bar{\psi}_{\delta t}) \cdot C_{\rm Ia}(\bar{\psi}_{\delta t})$, i.e., the overall SNIa realisation parameter. We set this parameter to be a constant in our `fiducial SNIa model'. It could become higher for more massive elliptical galaxies because the stars are formed on average in a more crowded and metal-rich environment. This leads to a higher fraction of hard binaries, larger asymptotic giant branch (AGB) star radii, and a more efficient accretion onto the pre-supernova white dwarf \citep{1996ApJ...470L..97H}. We note that the possible variation of the overall SNIa realisation parameter could be due to a variation of both SFR and metallicity but these two dependencies degenerate in the presented description (Eq.~\ref{eq:N_Ia} and Eq.~\ref{eq:SNIa_renormalization} below) which is only a function of SFR because both the peak SFR and mean stellar metallicity of E~galaxies depend on their mass monotonically.


In order to calibrate the overall SNIa realisation parameter $B_{\rm bin}(\bar{\psi}_{\delta t}) \cdot C_{\rm Ia}(\bar{\psi}_{\delta t})$, we start from a given galaxy-wide 10~Myr averaged SFR, $\psi_0=1 M_\odot$/yr. The value of $B_{\rm bin}(\psi_0) \cdot C_{\rm Ia}(\psi_0)$ can be determined by the observed number of SNIa events in nearby stellar systems assuming they have the canonical gwIMF and a Galactic SFR of $\psi_0$. The local SNIa incidence per unit mass of stars formed is determined by the observation of different systems with different SFR. Therefore, the calibration assuming a mean SFR of the local universe of $\psi_0=1 M_\odot$/yr may not be exactly correct but it has led to successful reproduction of the work of other groups \citep{2019A&A...632A.110Y,2020A&A...637A..68Y}. Following \citet{2012PASA...29..447M}, we set
\begin{equation}\label{eq:normalization}
    n_{\rm Ia}(t=10~\mathrm{Gyr}, \xi_{\mathrm{canonical}}, \psi_0)=0.0022/M_\odot,
\end{equation}
where $\xi_{\mathrm{canonical}}$ denotes that the relation holds when adopting the canonical Kroupa IMF (\citealt{2001MNRAS.322..231K}, i.e. Eq.~\ref{eq:IMF} with $\alpha_1=1.3$ and $\alpha_2=\alpha_3=2.3$).

\begin{figure}
    \centering
    \includegraphics[width=\hsize]{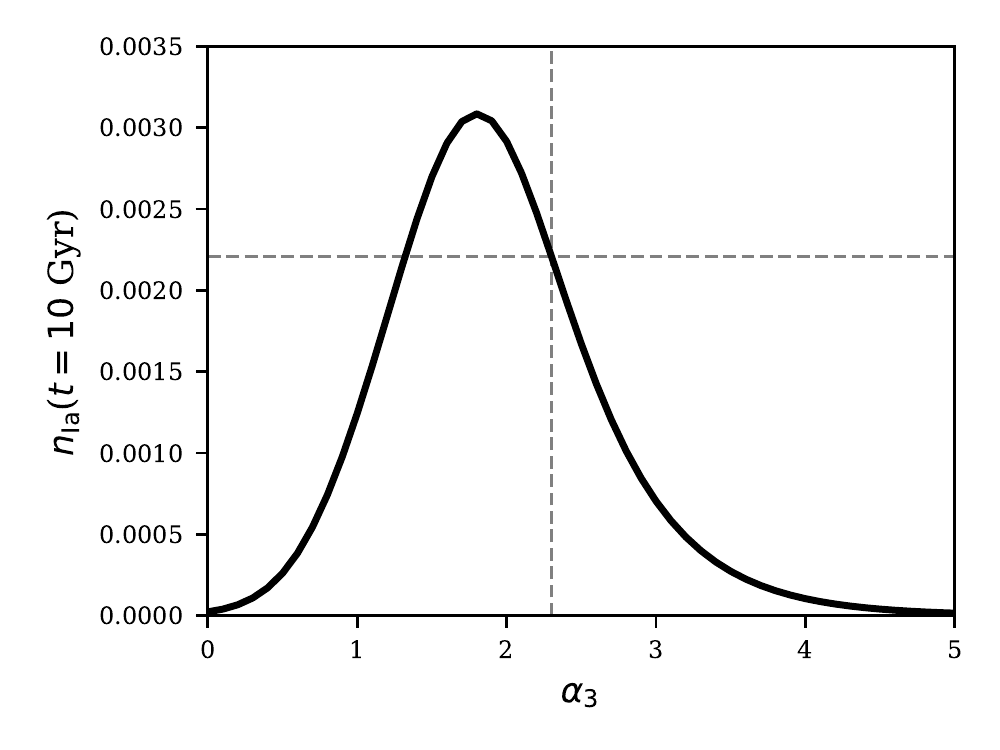}
    \caption{The total number of SNIa explosions for a SSP after 10 Gyr of the birth of the SSP per unit stellar mass of the SSP predicted by Eq.~\ref{eq:SNIa rate} to \ref{eq:normalization}, assuming constant parameter $B_{\rm bin}$ and $C_{\rm Ia}$ and an IMF defined in Eq.~\ref{eq:IMF}, with $\alpha_1=1.3$, $\alpha_2=2.3$, as a function of $\alpha_3$. A smaller $\alpha_3$ indicates a flatter IMF/gwIMF for the massive stars, i.e. a top-heavy IMF/gwIMF. Very large $\alpha_3$ mean that essentially no stars above $1 M_\odot$ are formed. The function is calibrated by the vertical and horizontal dashed lines indicating the canonical IMF/gwIMF slope, $\alpha_3=2.3$, and the value $n_{\rm Ia}(t=10~\mathrm{Gyr}, \xi_{\mathrm{canonical}})=0.0022/M_\odot$ suggested by \citet{2012PASA...29..447M} for the local universe, respectively.}
    \label{fig:SNIa_variation}
\end{figure}

Under the fiducial SNIa model with a constant overall SNIa realisation parameter, $B_{\rm bin}(\bar{\psi}_{\delta t}) \cdot C_{\rm Ia}(\bar{\psi}_{\delta t}) = B_{\rm bin}(\psi_0) \cdot C_{\rm Ia}(\psi_0)$, the above formulation in Eq.~\ref{eq:N_Ia} is completely defined and quantifies how $n_{\rm Ia}(t=10~\mathrm{Gyr}, \xi)$ varies given different gwIMFs, $\xi(m)$, and SFRs, $\bar{\psi}_{\delta t}$. Note that we drop $\bar{\psi}_{\delta t}$ in $n_{\rm Ia}(t, \xi, \bar{\psi}_{\delta t})$ because this dependency is contained in the gwIMF, $\xi$, by the latter being a function of $\bar{\psi}_{\delta t}$. As an example, Fig.~\ref{fig:SNIa_variation} demonstrates how $n_{\rm Ia}(t=10~\mathrm{Gyr}, \xi)$ varies with the IMF power-law index, $\alpha_3$. As can be seen, the SNIa production efficiency is maximised for $\alpha_3 \approx 1.8$ and suppressed for extremely top-heavy IMF cases. Such stellar populations may not be impossible: $\alpha_3<1$ for young globular clusters with masses larger than about $10^8 M_\odot$ according to \citet[their fig. 2]{2012MNRAS.422.2246M}, while $\alpha_3>4$ for the gwIMF in dwarf galaxies with $\bar{\psi}_{\delta t}<10^{-4} M_\odot$/yr according to \citet[their fig. 6]{2017A&A...607A.126Y}.

\begin{figure}
    \centering
    \includegraphics[width=\hsize]{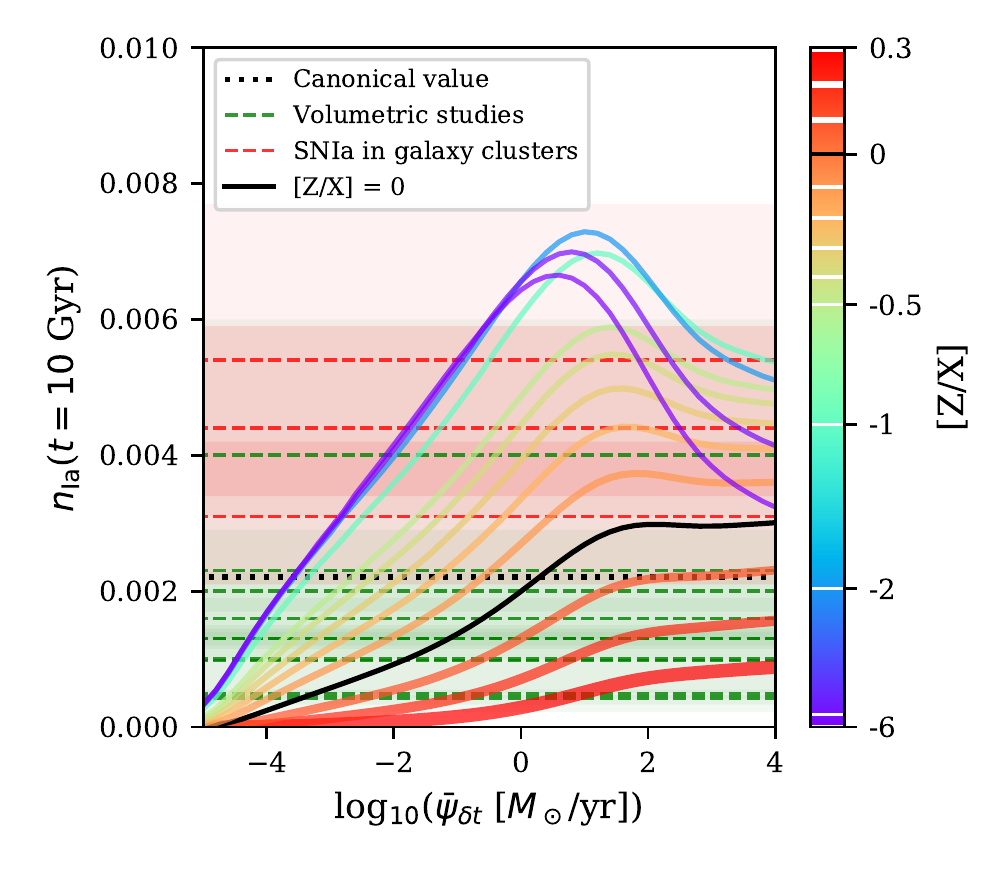}
    \caption{The 10 Gyr time-integrated number of SNIa per unit stellar mass formed for the fiducial SNIa model calculated according to Eqs~\ref{eq:SNIa rate} to \ref{eq:normalization}. The fiducial SNIa model assumes a constant overall SNIa realisation parameter and the gwIMF as given by the IGIMF theory (Eq.~\ref{eq:IGIMF}) as a function of the galaxy-wide SFR, $\bar{\psi}_{\delta t}$, and metallicity, [Z/X]. The black line is the relation for $[Z/X]=0$. Other lines with different colours represent different values of $[Z/X]$ as indicated by the white stripes on the colour-map on the right, being $[Z/X]=0.3$, 0.2, 0.1, -0.1, -0.2, -0.3, -0.4, -0.5, -1, -2, -4, and -6. The black horizontal dotted line represents the canonical $n_{\rm Ia}(t=10~\mathrm{Gyr}, \xi_{\mathrm{canonical}})$ value of $0.0022/M_\odot$ \citep{2012PASA...29..447M}. The green and red horizontal dashed lines indicate observational constraints on the $n_{\rm Ia}(t=10~\mathrm{Gyr})$ for SNIa surveys up to a certain redshift and in galaxy clusters, respectively. The shaded regions represent the uncertainty ranges of the horizontal dashed lines. References are given in Table~\ref{tab:Observed SNIa rate}.
    }
    \label{fig:SNIa_variation_IGIMF}
\end{figure}

Since the gwIMF predicted by the IGIMF theory is a function of the galaxy-wide SFR and metallicity (see Eq.~\ref{eq:IGIMF}), Fig.~\ref{fig:SNIa_variation_IGIMF} shows how $n_{\rm Ia}(t=10~\mathrm{Gyr})$ changes as a function of these two parameters for our fiducial SNIa model. The resulting $n_{\rm Ia}(t=10~\mathrm{Gyr})$ for the metal-poor galaxies can be much higher than the values shown in Fig.~\ref{fig:SNIa_variation} because they have a bottom-light gwIMF according to Eq.~\ref{eq:alpha18}.
The galaxy-wide SFR also affects the number of SNIa per stellar mass formed because of Eq.~\ref{eq:alpha3}. The calculated $n_{\rm Ia}(t=10~\mathrm{Gyr})$ variation due to different gwIMF agrees with the observational constraints for SNIa surveys (listed in Table~\ref{tab:Observed SNIa rate}) and could be a natural explanation for the difference in these observations. This is not the case for the older version of the IGIMF formulations summarised in \citet[their table 3]{2018A&A...620A..39J} as is demonstrated by Fig.~\ref{fig:SNIa_variation_IGIMF1} to \ref{fig:SNIa_variation_IGIMF3} in the Appendix.
\begin{table}
\caption{Observational estimations on the time-integrated number of SNIa per stellar mass formed.}
\label{tab:Observed SNIa rate}
\centering
\begin{tabular}{c c}
\hline\hline
$N_{\rm Ia}\cdot 10^3~[M_\odot^{-1}]$ & Reference \\
\hline
\multicolumn{2}{c}{Volumetric and field galaxies} \\
\hline
$2.3\pm0.6$ & \citet{2011MNRAS.412.1508M} \\
$1.3\pm0.15$ & \citet{2012MNRAS.426.3282M} \\
$0.485\pm0.065$ & \citet{2012AJ....144...59P} \\
$0.43_{-0.1}^{+0.04}$ & \citet{2013MNRAS.430.1746G} \\
$0.98_{-0.76}^{+1.3}$ & \citet{2014AJ....148...13R} \\
$1.3\pm0.1$ & \citet{2017ApJ...848...25M} \\
$1.6\pm0.3$ & \citet{2017ApJ...848...25M}  \\
$2\pm1$ & \citet{2019ApJ...882...52H} \\
$4_{-1}^{+2}$ & \citet{2019ApJ...882...52H} \\
\hline
\multicolumn{2}{c}{Galaxy clusters} \\
\hline
$4.4_{-1}^{+1.5}$ & \citet{2010ApJ...722.1879M} \\
$5.4\pm2.3$ & \citet{2017ApJ...848...25M} \\
$3.1_{-1.0}^{+1.1}$ & \citet{2021MNRAS.502.5882F} \\
\hline
\end{tabular}
\end{table}

In addition, we divide the estimations of the SNIa production efficiency in Table~\ref{tab:Observed SNIa rate} into two groups, those targeted at galaxy clusters and volume-limited and field galaxies (plotted as the red and green shaded regions, respectively, in Fig.~\ref{fig:SNIa_variation_IGIMF}).
These results suggest that galaxy clusters have a higher SNIa production efficiency. The $n_{\rm Ia}(t=10~\mathrm{Gyr})$ of our predicted high-SFR metal-rich galaxies is lower than these estimates. As mentioned above, it is reasonable to consider that the overall SNIa realisation parameter is higher for the galaxies that have experienced more intense star formation in a galaxy cluster environment. We discuss such scenarios in Section~\ref{sec:SFT with variable SNIa rate} below.

\subsection{The galaxy chemical evolution model}\label{sec:chemical model}

Motivated by the monolithic collapse formation scenario of E~galaxies (see Section~\ref{sec:intro}) and following \citet{1999MNRAS.302..537T}, the single-zone closed-box galaxy chemical evolution modelling computer programme, GalIMF, described in \citet{2019A&A...629A..93Y} is applied.
GalIMF assumes that all the gas is always well mixed (i.e. single-zone) and that the stars inject a certain amount of mass of each element into the gas when they exhaust their lifetime (non-instantaneous recycling of metals). The smallest time-step in our galaxy-wide calculation is $\delta t = 10\,$Myr because this is the timescale required for a galaxy to fully populate the ECMF \citep{2017A&A...607A.126Y} and the gwIMF cannot be defined over shorter timescales (see also Section~\ref{sec:IGIMF}).

The stellar yield table assumed here is the same as in \citet{2020A&A...637A..68Y}. 
That is, the stellar lifetime, remnant mass, and metal yield for low-mass stars are adopted from \citet{2001A&A...370..194M} and for massive stars from \citet{2006ApJ...653.1145K}. The stellar yield is given according to the initial mass and metallicity of a star (cf. \citealt[their fig. 3 and 4]{2019A&A...629A..93Y}). The yield table is first interpolated in the dimension of stellar-mass while for stars with a mass higher or lower than the mass range provided in the table, a fixed value is applied being, respectively, the value for the most massive or least-massive star (cf. \citealt[their fig. 8]{2020A&A...637A..68Y}). Then the table is interpolated again in the dimension of the stellar initial metallicity. For metallicities higher/lower than the range provided in the table, the value from the largest/lowest metallicity in the table is applied, respectively.
We note that the yield table for massive stars is different from our previous work on constraining the SFT of elliptical galaxies \citep{2019A&A...632A.110Y} which follows the prescription of \citet{1999MNRAS.302..537T}. This introduces a systematic difference but preserves the general trend of the $\tau_{\rm SF}$--$M_{\rm dyn}$ relation and does not change our conclusions, as is explained in Section~\ref{sec:Stellar yield}.

\section{Observational constrains}\label{sec:Observational constrain}

\begin{figure}
    \centering
    \includegraphics[width=\hsize]{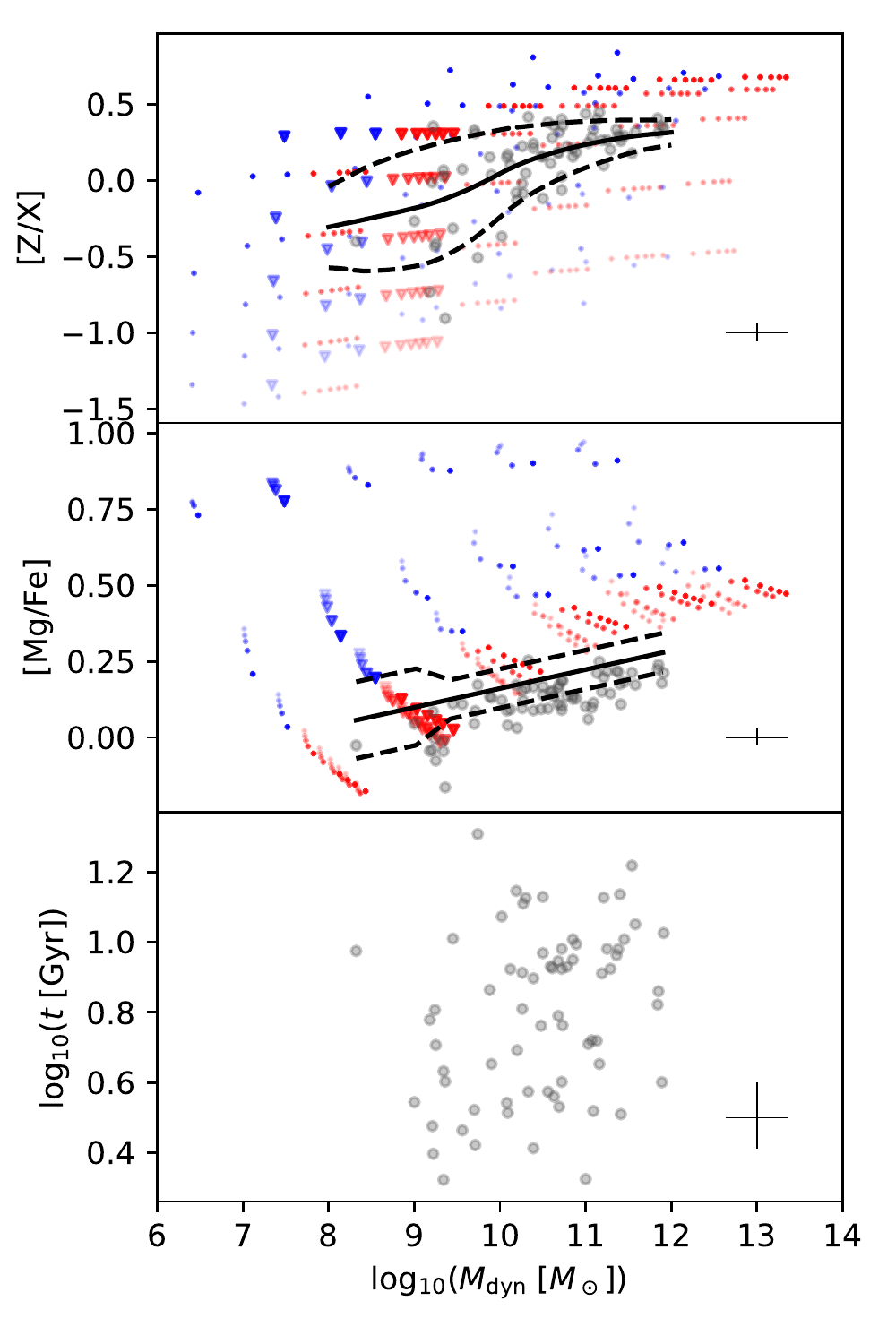}
    \caption{The [Z/X], [Mg/Fe], and age of the E galaxies as a function of their dynamical masses. The grey circles are observed galaxies from \citet{2010MNRAS.402..173A} with typical uncertainty shown at the lower right corner of each panel. The solid and dashed lines are, respectively, the mean and standard deviation of the observational constraints. The lines in the upper panel are given by \citet{2010MNRAS.402..173A} and smoothed by \citet{2019A&A...632A.110Y}. The mean relation for the middle panel follows \citet{2005ApJ...621..673T} while the standard deviation of the relation follows \citet{2016ApJ...829L..26L}. The points and triangles are the resulting $M_{\rm dyn}$, [Z/X], and [Mg/Fe] values of our fiducial galaxy models adopting the IGIMF theory and a constant overall SNIa realisation parameter defined in Section~\ref{sec:SNIa}. The triangles are models with $\mathrm{log}_{10}(\bar{\psi}_{\delta t})=0$ while the points are models with other SFRs. The red and the blue dots are models with a SFT $\tau_{\rm SF}\geq1$ Gyr and $\tau_{\rm SF}<1$ Gyr, respectively. The darker enhanced colour are models with a higher $g_{\rm conv}$ (defined in Section~\ref{sec:Method}).}
    \label{fig:data}
\end{figure}

The galaxy models need to be compared with the observations to constrain the parameters of the galaxy models such as the SFTs. Here we apply the observational $[\mathrm{Z/X}]$ and $[\mathrm{Mg/Fe}]$ constrains shown by the thick solid and dashed lines in Fig.~\ref{fig:data}. 
The $[\mathrm{Z/X}]$--$M_{\rm dyn}$ relation is constrained by \citet{2010MNRAS.402..173A} and the mean and the standard deviation of the relation is calculated using the procedure detailed in \citet[their section 2.1]{2019A&A...632A.110Y}. The comparison between the galaxy dynamical mass derived from observation and the dynamical mass from our model is detailed in \citet[their section 2.3]{2019A&A...632A.110Y}.

We apply the [Mg/Fe]--$M_{\rm dyn}$ relation given by \citet[their eq. 3]{2005ApJ...621..673T} but a larger [Mg/Fe] scatter for galaxies with a mass smaller than $10^{9.218} M_\odot$ as is suggested by \citet[their table 1]{2016ApJ...829L..26L}, calculated with the galaxy-mass--central-velocity-dispersion relation given using \citet[their eq. 2]{2005ApJ...621..673T}. 
The [Mg/Fe]--$M_{\rm dyn}$ relation given by \citet{2005ApJ...621..673T} has a higher [Mg/Fe] value of about 0.05 dex than the relation given by the galaxies in \citet{2010MNRAS.402..173A} their table B1 (see the solid line and the grey circles in Fig.~\ref{fig:data}). We adopt the \citet{2005ApJ...621..673T} relation, because it is supported by the more recent and complete study by \citet{2016ApJ...829L..26L} and this choice is consistent with our previous study \citep{2019A&A...632A.110Y}.
The standard deviation of [Mg/Fe], $\sigma_{\rm Mg/Fe}$, is 0.126 for galaxies with $M_{\rm dyn}<10^{9.018} M_\odot$ and 0.064 for galaxies with $M_{\rm dyn}>10^{9.418} M_\odot$ and a linear transition in between.

The age of the observed galaxies given in \citet[their table B1]{2010MNRAS.402..173A} has a large scatter, $6.3^{+4.7}_{-2.7}$ Gyr. There is no strong correlation between the age and mass or abundance of the galaxies as is shown by the lower panel of Fig.~\ref{fig:data} although \citet{2005ApJ...621..673T} calculated that massive ETGs and ETGs in high-density environments form earlier statistically, a.k.a., `archaeological downsizing'. We neglect the age difference of these galaxies and evolve all our modelled galaxies to 14 Gyr for simplicity. Assuming a different galaxy age for each galaxy or all the galaxies would not affect our conclusions because the modelled galaxies are assumed to have quenched their star formation completely. The properties of their stellar population would not have any significant change after the quenching.

These observational relations are derived from the ETG database but since elliptical and lenticular galaxies follow similar $[\mathrm{Z/X}]$--$M_{\rm dyn}$ and $[\mathrm{Mg/Fe}]$--$M_{\rm dyn}$ relations (cf. \citealt{2005MNRAS.362...41G,2010ApJ...722..491Z,2021MNRAS.504.3478C}) the difference of galaxy type is unlikely to affect our conclusions, although a follow-up study distinguishing galaxy types would be valuable.


\section{Method}\label{sec:Method}

Here we explain how the GalIMF code is used to constrain the SFTs of galaxies following the same method as in \citet{2019A&A...632A.110Y}. In short, there are three input parameters that describe the gas supply and SFH of a galaxy and there are three outputs after running the code that describe the mass and metal abundance of this galaxy after 14 Gyr. The output is compared with the observational values to evaluate the likelihood of the input parameters.

The chemical evolution of a set of different galaxies with different combinations of input parameters is calculated. Following \citet{2019A&A...629A..93Y} and \citet{2019A&A...632A.110Y}, the SFHs of the E~galaxies are assumed to have a constant SFR, $\bar{\psi}_{\delta t}$, over a certain SFT, $\tau_{\rm SF}$. A low level of star formation has been found common in "quenched" galaxies \citep{2020MNRAS.498.1002D}. This could be the cause of the large scatter of the observed abundance but does not have a significant effect on our results because we focus on the smoothed mean relation of galaxies. In addition, \citet{2020NatAs...4..252S} find only a percent of stellar mass to have been added to ETGs during the past 2 Gyr (see also \citealt{2020MNRAS.498.5652K} for an in-depth discussion of ETG assembly in connection to the formation of supermassive black holes). The third input parameter is the gas-conversion parameter, $g_{\rm conv}$, defined as the ratio between the accumulated stellar mass (total mass of stars ever formed in a galaxy) and initial-gas mass (total gas supply since the model assumes no galactic gas in-/out-flow). By adjusting the amount of initial gas supply for the same SFH, different $g_{\rm conv}$ values are tested. The input parameters for our fiducial SNIa model take the values of: $g_{\rm conv}=$ 0.05, 0.1, 0.2, 0.4, or 0.8, $\mathrm{log}_{10}(\bar{\psi}_{\delta t}[M_\odot/\mathrm{yr}])=$ -1.0, 0, 1.0, ..., or 4.0, and $\tau_{\rm SF}=$ 0.05, 0.2, 0.5, 0.75, 1, 1.25, 1.5, 1.75, 2, 2.25, 2.5, 2.75, 3, 3.5, or 4 [Gyr]. In total, $5\times6\times15=450$ galaxy models are calculated. The other models discussed in Section~\ref{sec:SFT with variable SNIa rate} are similar but focus on different parameter ranges.

All galaxies are evolved to 14 Gyr for comparison with the observations because the evolution of the mean stellar properties later than 1 Gyr after cessation of star formation for a galaxy is negligible. Therefore, the applied timesteps for each model are 0.01, 0.02, ..., $\tau_{\rm SF}-0.01$, $\tau_{\rm SF}$, and 14 [Gyr].
The resulting galactic stellar plus remnant mass, $M_{\rm dyn}$, mean stellar metallicity, and mean stellar [Mg/Fe] for the model galaxies with every combination of the input parameters are calculated and plotted in Fig.~\ref{fig:data}. Then the results are interpolated in 3D for the three parameters ($g_{\rm conv}$, $\bar{\psi}_{\delta t}$, and $\tau_{\rm SF}$) and compared with the mean observational values of galaxies with the same mass.

The goodness of the fit for each galaxy model is calculated as
\begin{equation}\label{eq:likelihood}
\begin{split}
& p(g_{\rm conv}, \tau_{\rm SF}) = p_{\rm Z/X} \times p_{\rm Mg/Fe},\\
& p_{\rm Z/X} = \\
& 1-\mathrm{erf}\left[\left([\mathrm{Z/X}]_{\rm obs}-[\mathrm{Z/X}]_{\rm mod}(g_{\rm conv}, \tau_{\rm SF})\right)/\sigma_{\mathrm{Z/X}}/\sqrt{2}\right], \\
& p_{\rm Mg/Fe} = \\
& 1-\mathrm{erf}\left[\left([\mathrm{Mg/Fe}]_{\rm obs}-[\mathrm{Mg/Fe}]_{\rm mod}(g_{\rm conv}, \tau_{\rm SF})\right)/\sigma_{\mathrm{Mg/Fe}}/\sqrt{2}\right],
\end{split}
\end{equation}
where $[\mathrm{Z/X}]_{\rm obs}$ and $[\mathrm{Mg/Fe}]_{\rm obs}$ are the mean observational values and the $\sigma$ stands for the standard deviation of the element abundance ratios for different galaxies for a given mass (see \citealt{2019A&A...632A.110Y}). For each $M_{\rm dyn}$ and $\tau_{\rm SF}$, the highest $p$ value with any $g_{\rm conv}$ is calculated and shown in Figs~\ref{fig:best_fit_SFT} and \ref{fig:best_fit_SFT_SNIa}. Then the $\tau_{\rm SF}$ with the highest possible $p$ for a given $M_{\rm dyn}$ indicates the mean SFT for the elliptical galaxies with the mass $M_{\rm dyn}$.

Finally, we modify the default assumptions in the GalIMF code, including the IMF and SNIa formulation to study their effects on the results. That is, we calculate a galaxy model grid for each set of IMF/SNIa assumptions.

\section{Results}\label{sec:Results}

In Section~\ref{sec: Results Canonical IMF} the results obtained by applying an invariant canonical IMF for all galaxies are documented. This serves as a bench-mark model-set for the more realistic case documented in Section~\ref{sec: results assuming IGIMF} where the galaxy-wide IMF changes with changing conditions as is calculable self-consistently (from the time-dependent SFR and metallicity) using the IGIMF theory.

\subsection{The canonical IMF}\label{sec: Results Canonical IMF}
\subsubsection{The canonical IMF and invariant SNIa production efficiency}
We have explored the possible $\tau_{\rm SF}$--$M_{\rm dyn}$ relations in \citet{2019A&A...632A.110Y} assuming the invariant canonical gwIMF. The results are plotted in Fig.~\ref{fig:best_fit_SFT} as dotted yellow lines along with the estimates given by \citet{2004MNRAS.347..968P} and \citet{2005ApJ...621..673T}.
The blue dotted line is from \citet{2005ApJ...621..673T}. It assumes a constant total-star-to-gas-mass fraction, $g_{\rm conv}$, in their galaxy chemical evolution model and takes into account only the [Mg/Fe] constraints. On the other hand, \citet[blue triangles]{2004MNRAS.347..968P} and \citet[yellow dotted lines]{2019A&A...632A.110Y} adopt a flexible total-star-to-gas-mass fraction and fit simultaneously the stellar [Mg/Fe] and the metallicity, resulting in steeper relations.

These results, constrained by the chemical abundance of the galaxies, suggest shorter SFTs than the timescale estimated by the SPS studies \citep{2011MNRAS.418L..74D,2015MNRAS.448.3484M}. As is mentioned in Section~\ref{sec:intro}, the cosmological hydrodynamical simulations of galaxy formation taking into account a realistic gas recycling time also have difficulty in reproducing the stellar [Mg/Fe] and metallicity of the most massive galaxies when assuming the canonical universal IMF. The short SFT for the most massive E~galaxies suggested by their high [Mg/Fe] value may be relaxed if they have a top-heavy IMF or there are fewer SNIa in these galaxies.

\subsubsection{The canonical IMF and variable SNIa production efficiency}

To have a longer SFT (more SNIa explosion and iron produced during the star formation) for the same observed [Mg/Fe] of massive ETGs, a lower SNIa production efficiency would be required assuming an invariant IMF. This scenario contradicts the expectation that massive galaxies that formed more massive clusters should have formed more hard WD--WD and WD--stellar binaries and, therefore, should have more SNIa explosions \citep{2002ApJ...571..830S}. This has led us to consider the observationally motivated variation of the gwIMF, as described in Section~\ref{sec: results assuming IGIMF}.




\subsection{The IGIMF theory}\label{sec: results assuming IGIMF}

It is always possible to fit the observed galactic abundance if one is allowed to fine-tune the gwIMF formulation with no constraints. However, one needs to follow the observed IMF variation of the resolved stellar populations, i.e., the gwIMF is not arbitrary. The here applied IGIMF formulation (Section~\ref{sec:IGIMF}) is determined by the empirically deduced IMF and ECMF variations and the calculated gwIMF automatically emerges and happens to be consistent with the observed gwIMF of vastly different systems.

\subsubsection{The IGIMF theory and the fiducial SNIa model}\label{sec:SFT with IGIMF}

For the fiducial SNIa model, we assume the parameters $B_{\rm bin}$ and $C_{\rm Ia}$ in Eq.~\ref{eq:N_Ia} to be invariant, i.e., SNIa production efficiency depends only on the gwIMF but not the star formation environment. Then, there is only the one unknown overall SNIa realisation parameter in Eq.~\ref{eq:SNIa rate}, i.e., $B_{\rm bin}(\psi_0) \cdot C_{\rm Ia}(\psi_0)$, which can be determined by Eq.~\ref{eq:normalization}. Under this assumption, the number of SNIa is only affected by the shape of the gwIMF (through the $\xi$ related parameters in Eq.~\ref{eq:N_Ia}) but not directly by the physical condition (density and metallicity) of the star formation cloud.

The resulting [Z/X]--$M_{\rm dyn}$ and [Mg/Fe]--$M_{\rm dyn}$ relations of our galaxy models adopting the IGIMF theory with different combinations of input $\bar{\psi}_{\delta t}$, $\tau_{\rm SF}$, and $g_{\rm conv}$ (detailed in Section~\ref{sec:Method}) are plotted in Fig.~\ref{fig:data}.
Comparing the results with different input parameters shown by the different symbols (see e.g. the middle panel, the 6 nearly vertical sequences of galaxy model results, from left to right, are models with $\mathrm{log}_{10}(\bar{\psi}_{\delta t})=$ -1.0, 0, 1.0, ..., or 4.0), it appears that the SFR is the main factor for establishing the resulting galaxy mass. Darker points have a higher $g_{\rm conv}$, i.e., more stellar mass formed per unit primordial gas provided. Thus, the $g_{\rm conv}$ value dominates the [Z/X] variation as is shown by the top panel. 
The change of SFT from 0.05 to 4 Gyr results in the galaxy model groups from the top to the bottom in the middle panel of Fig.~\ref{fig:data}, where the models with $\tau_{\rm SF}\geq1~\mathrm{Gyr}$ are plotted in red symbols, and this change dominates the [Mg/Fe] variation.

\begin{figure}
    \centering
    \includegraphics[width=\hsize]{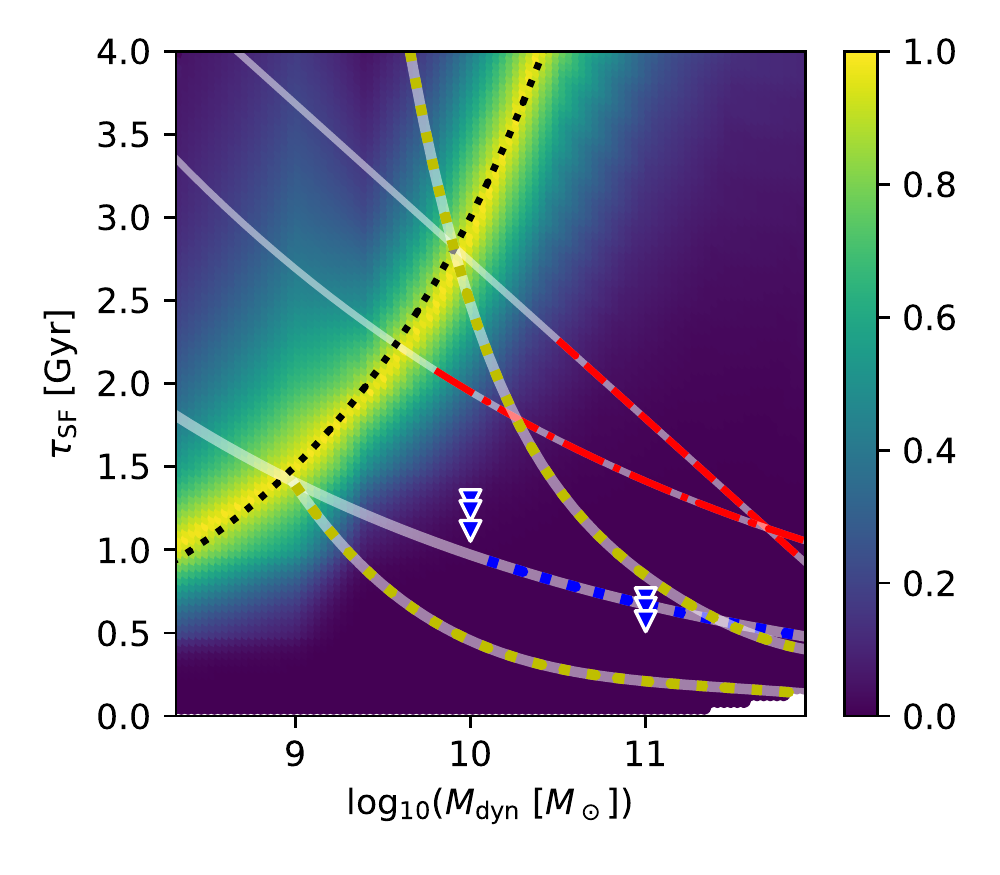}
    \caption{$\tau_{\rm SF}$ as a function of $M_{\rm dyn}$. The colour bar indicates the goodness of the fit, $p(g_{\rm conv}, \tau_{\rm SF})$ (Eq.~\ref{eq:likelihood}) for our fiducial SNIa model. For a given mass and SFT, $\tau_{\rm SF}$, any star-to-gas-mass fraction, $g_{\rm conv}$, is allowed. The most likely SFT for a given galaxy mass is indicated by the yellow ridgeline. The black dotted curve (Eq.~\ref{eq:SFT fit}) is a power-law fit of the yellow ridgeline. The red dashed curve and the red dash-dotted curve (Eq.~\ref{eq:SFT McDermid}) are the SFT constrained by SPS studies by \citet{2011MNRAS.418L..74D} and \citet{2015MNRAS.448.3484M}, respectively. As is introduced in Section~\ref{sec:intro}, The SFT suggested by chemical evolution studies assuming the invariant canonical gwIMF performed by \citet{2004MNRAS.347..968P} and \citet{2005ApJ...621..673T} are shown by the blue triangles and blue dotted curve, respectively. The lower and upper yellow dotted curves correspond, respectively, to polynomial regressions of the yellow ridgelines in figure~5 and~7 of \citet{2019A&A...632A.110Y}, derived using invariant canonical gwIMF. The mass ranges of the coloured curves are limited by the mass ranges of the galaxy data sets they are based on.}
    \label{fig:best_fit_SFT}
\end{figure}

The likelihoods of different SFTs for each galaxy mass are calculated according to Eq.~\ref{eq:likelihood} and shown in Fig.~\ref{fig:best_fit_SFT} by the colour map.
The best-fit SFT values (shown as the yellow ridgeline) are longer for more massive galaxies because they have top-heavier gwIMFs that lead to a larger production of magnesium.
A power-law function representing the yellow ridgeline in Fig.~\ref{fig:best_fit_SFT}, show by the black dotted line, is given by
\begin{equation}\label{eq:SFT fit}
    \tau_{\rm SF}~\mathrm{[Gyr]} = 0.003\cdot (M_{\rm dyn}/M_\odot)^{0.3},
\end{equation}
for a set of galaxies with a $M_{\rm dyn}$ above $10^{8.3}~M_\odot$, although there is only one galaxy in the data set with a mass below $10^{9}~M_\odot$ (see Fig.~\ref{fig:data}).
This resulting relation is not consistent with the relations suggested by the independent SPS studies shown by the red dashed and red dash-dot lines given by \citet{2011MNRAS.418L..74D} and \citet{2015MNRAS.448.3484M}, respectively. For example, in combination with eq.~3 of \citet{2005ApJ...621..673T}, \citet[their eq. 3]{2015MNRAS.448.3484M} suggests
\begin{equation}\label{eq:SFT McDermid}
    \tau_{\rm SF}~\mathrm{[Gyr]} = 49\cdot (M_{\rm dyn}/M_\odot)^{-0.14},
\end{equation}
for a set of galaxies with a $M_{\rm dyn}$ between about $10^{9.8}$ and $10^{12}~M_\odot$. This relation is shown as the red dash-dotted curve in Fig.~\ref{fig:best_fit_SFT}, \ref{fig:best_fit_SFT_SNIa}, and \ref{fig:best_fit}.
However, as shown next, taking into account that more massive E~galaxies are likely to have formed more massive and dense star-burst clusters which produce a larger fraction of SNIa-progenitor binaries per star \citep{2002ApJ...571..830S}, it is possible to obtain consistent results.


\subsubsection{The IGIMF theory and a boosted SNIa production efficiency for high-SFR galaxies}\label{sec:SFT with variable SNIa rate}

We define an SNIa realisation renormalization parameter,
$\kappa_{\rm Ia}(\bar{\psi}_{\delta t})$, which represents the variation of 
the overall SNIa realisation parameter (the terms independent from the IMF in Eq.~\ref{eq:N_Ia}) as a function of the galaxy-wide SFR,
\begin{equation}\label{eq:SNIa_renormalization}
    \kappa_{\rm Ia}(\bar{\psi}_{\delta t}) = \frac{B_{\rm bin}(\bar{\psi}_{\delta t}) \cdot C_{\rm Ia}(\bar{\psi}_{\delta t})}{B_{\rm bin}(\psi_0) \cdot C_{\rm Ia}(\psi_0)},
\end{equation}
where the constant $B_{\rm bin}(\psi_0) \cdot C_{\rm Ia}(\psi_0)$ is calibrated by reproducing this value for a single stellar population with the canonical \citet{2001MNRAS.322..231K} IMF while the variable $B_{\rm bin}(\bar{\psi}_{\delta t}) \cdot C_{\rm Ia}(\bar{\psi}_{\delta t})$ can become larger in massive galaxies, as mentioned in Section~\ref{sec:SNIa}.

\begin{figure}
    \centering
    \includegraphics[width=\hsize]{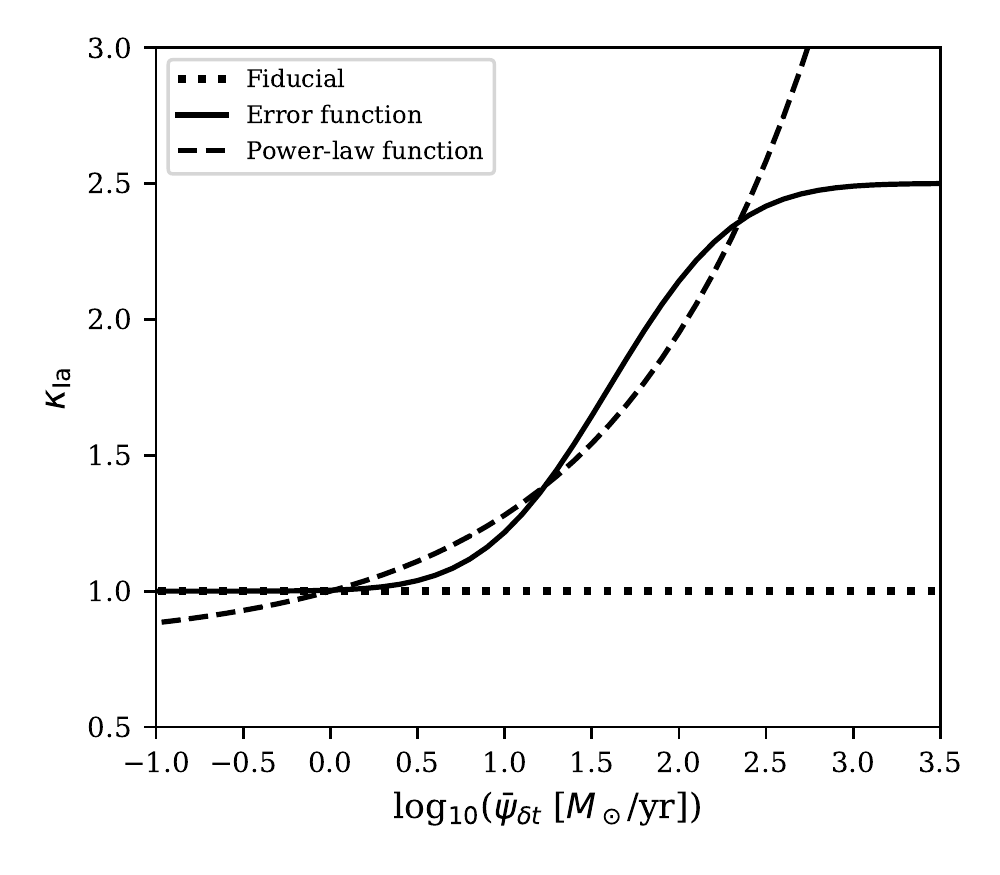}
    \caption{The SNIa realisation renormalization parameter, $\kappa_{\rm Ia}$ defined in Eq.~\ref{eq:SNIa_renormalization}, as a function of the galaxy-wide SFR, $\bar{\psi}_{\delta t}$. Three different models are tested and the corresponding best fit SFT--galaxy-mass relations are shown in Fig.~\ref{fig:best_fit_SFT}, \ref{fig:best_fit_SFT_SNIa}, and \ref{fig:SFT_different_k_SFR_relations} for the fiducial model (dotted line), error function model (solid line), and power-law function model (dashed line), respectively. The error function model, formulated in Eq.~\ref{eq:error function}, best reproduces the observed ETGs.
    }
    \label{fig:SNIa_renormalization}
\end{figure}

\begin{figure}
    \centering
    \includegraphics[width=\hsize]{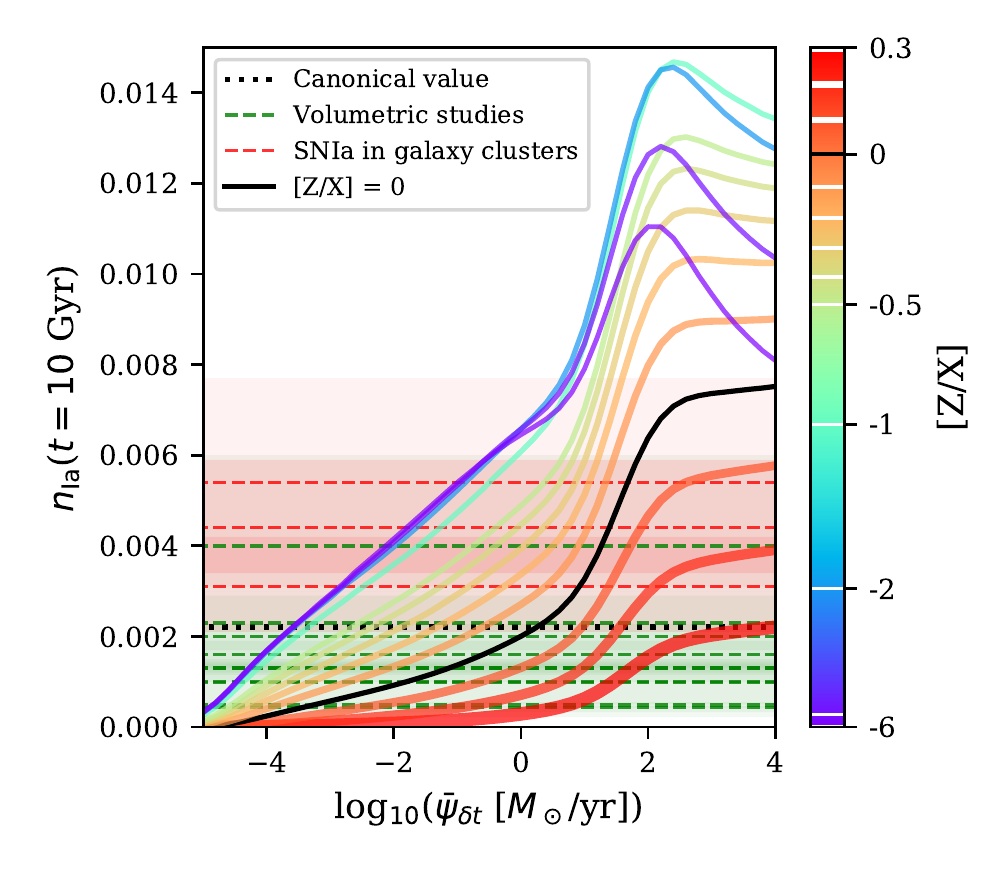}
    \caption{The 10 Gyr time-integrated number of SNIa per unit stellar mass formed for the error function model. Same as Fig.~\ref{fig:SNIa_variation_IGIMF} but here the calculation assumes the variable $\kappa_{\rm Ia}$--$\bar{\psi}_{\delta t}$ relation defined by Eq.~\ref{eq:error function} and shown by the solid curve in Fig.~\ref{fig:SNIa_renormalization} instead of a constant overall SNIa realisation parameter (the dotted line in Fig.~\ref{fig:SNIa_renormalization}).}
    \label{fig:SNIa_variation_IGIMF_varKappa}
\end{figure}

\begin{figure}
    \centering
    \includegraphics[width=\hsize]{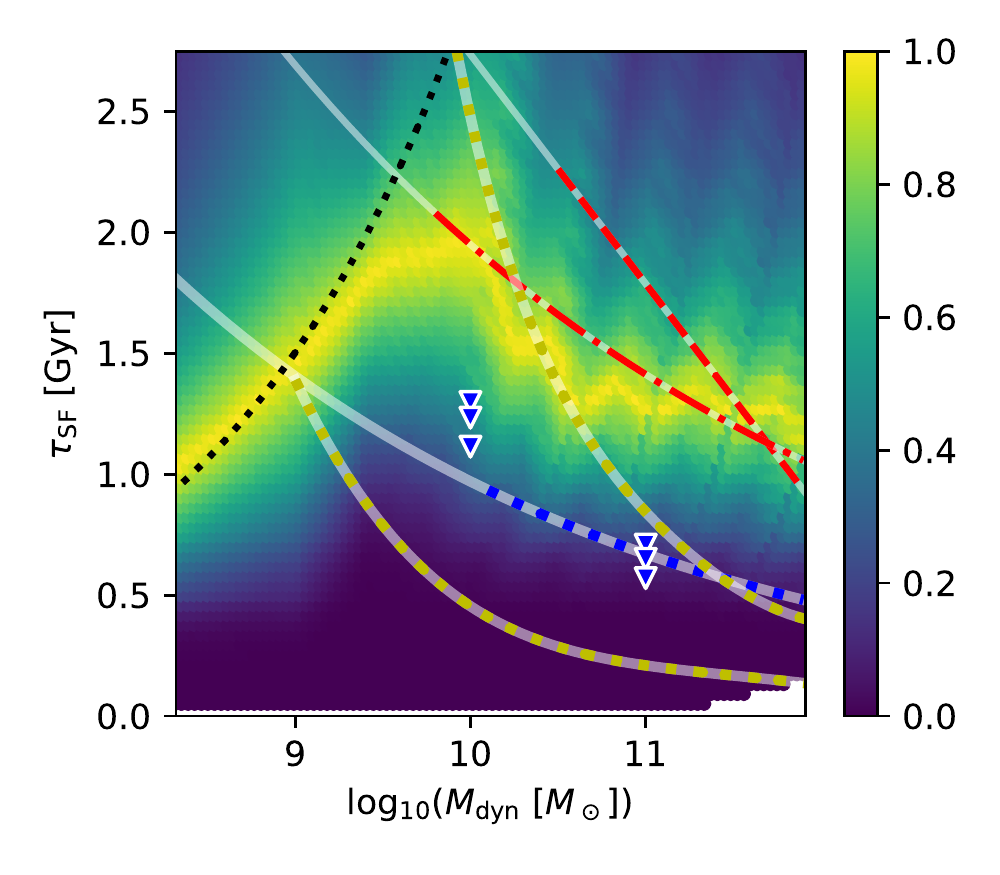}
    \caption{$\tau_{\rm SF}$ as a function of $M_{\rm dyn}$ for the error function model. Same as Fig.~\ref{fig:best_fit_SFT} but the results shown by the colour map adopt the variable $\kappa_{\rm Ia}(\bar{\psi}_{\delta t})$ defined by Eq.~\ref{eq:error function} and shown by the solid curve in Fig.~\ref{fig:SNIa_renormalization}. The best-fit solutions (the yellow ridge line) for different $M_{\rm dyn}$ can be described by two relations, i.e., the black dotted curve (Eq.~\ref{eq:SFT fit}) for $M_{\rm dyn}<10^{9.5}~M_\odot$ and the red dash-dotted curve (Eq.~\ref{eq:SFT McDermid}) for $M_{\rm dyn}>10^{9.8}~M_\odot$.}
    \label{fig:best_fit_SFT_SNIa}
\end{figure}

Through trial and error, we find that assuming the error function
\begin{equation}\label{eq:error function}
    \kappa_{\rm Ia}(\bar{\psi}_{\delta t}) = 1.75 + 0.75 \cdot \mathrm{erf}[\mathrm{log}_{10}(\bar{\psi}_{\delta t}) \cdot 1.25 - 2],
\end{equation}
where erf stands for the Gauss error function (the solid curve in Fig.~\ref{fig:SNIa_renormalization}) leads to a result (the yellow ridgeline in Fig.~\ref{fig:best_fit_SFT_SNIa}) that roughly fits the $\tau_{\rm SF, SPS}$--$M_{\rm dyn}$ relation suggested by \citet{2015MNRAS.448.3484M}. The SNIa production efficiency is boosted by $\kappa=2.5$ for the most massive ellipticals compared to the unchanged value of $\kappa=1$ for the low-mass galaxies. This increases the number of SNIa per stellar mass formed for high-SFR galaxies as is shown in Fig.~\ref{fig:SNIa_variation_IGIMF_varKappa} in comparison with Fig.~\ref{fig:SNIa_variation_IGIMF}, and therefore, it decreases their SFT estimation. The significantly increased SNIa production efficiency of the high-SFR galaxies shown in Fig.~\ref{fig:SNIa_variation_IGIMF_varKappa} are consistent with the observational findings that the SNIa production efficiency in galaxy clusters (red regions in Fig.~\ref{fig:SNIa_variation_IGIMF_varKappa}), which host massive ETGs, is only 2 to 3 times higher than for field galaxies (green regions in Fig.~\ref{fig:SNIa_variation_IGIMF_varKappa}, see e.g. \citealt{2018MNRAS.479.3563F,2021MNRAS.502.5882F}) because the high-SFR galaxies have an overall metal-rich stellar population, leveraging down the $n_{\rm Ia}(t=10~\mathrm{Gyr})$ value.

Fig.~\ref{fig:best_fit_SFT_SNIa} shows the resulting $\tau_{\rm SF}$--$M_{\rm dyn}$ relation for the error function model. The E~galaxies with mass smaller than about $10^{9.5} M_\odot$ have a most-likely SFT that is shorter for lower-mass galaxies, following the same black dotted curve as in Fig.~\ref{fig:best_fit_SFT}, i.e., Eq.~\ref{eq:SFT fit}. This could arise because low-mass galaxies have a shallower potential and lose their gas supply more easily. Although the SFT for individual dwarf ellipticals can have a large scatter (cf. Fig.~\ref{fig:data}), our result is in good agreement with the independent SFT estimations for the ultra-diffuse galaxy Dragonfly~44 \citep{2019ApJ...884L..25H} using the SPS method and for the ultra-faint dwarf galaxy Bo\"otes~I \citep{2020A&A...637A..68Y} analysing the chemical enrichment history of the galaxy, both finding a SFT less than 1 Gyr. On the other hand, E~galaxies with masses larger than about $10^{9.5} M_\odot$ have SFT values that are shorter with increasing $M_{\rm dyn}$, following the relation suggested by \citet{2015MNRAS.448.3484M}, i.e., Eq.~\ref{eq:SFT McDermid}. These massive galaxies are likely regulated by the gas collapsing timescale. As a result, the highest SFT is reached for galaxies with about $10^{9.5} M_\odot$. We note that the galaxy mass reaching the peak SFT can be smaller if $\kappa$ is lower than 1 for low-SFR galaxies.
It is possible to further fine-tune the $\kappa_{\rm Ia}(\bar{\psi}_{\delta t})$ function to adjust the resulting $\tau_{\rm SF}$--$M_{\rm dyn}$ relation, which we do not do considering the substantial computational cost. The short SFT for low-mass galaxies is concluded here due to a top-light gwIMF for the low-SFR galaxies, in distinct difference from the conclusions reached by \citet{2005ApJ...621..673T} and \citet{2019A&A...632A.110Y} assuming the invariant canonical IMF and the closed-box chemical evolution model. However, the very top-light gwIMF of the present-day low-SFR galaxies (\citealt{2009ApJ...706..599L}, \citealt{2009MNRAS.395..394P}, and \citealt[their fig. 6]{2017A&A...607A.126Y}) may not represent the gwIMF of dwarf galaxies during their formation. Dwarf galaxies can be perturbed and have a more complex SFH than the monolithic collapse scenario assumed here (see discussions in Section~\ref{sec:Closed-box model}). With a fluctuated SFR, the stellar population of the dwarf galaxies are formed with a higher instantaneous SFR, $\bar{\psi}_{\delta t}$, than the mean SFR over the same SFT, leading to a less top-light gwIMF and a longer derived SFT (cf. Section~\ref{sec:Caveat on SPS}). In other words, for an isolated E~galaxy, the difference between its real SFT and the calculated SFT using its abundance constraints and a constant SFR assumption indicates how fluctuated its SFR is.

In addition, we explore the possibility that the SNIa production efficiency for high SFR galaxies continuous to increase following a power-law relation. In the case of the power-law function model shown by the dashed curve in Fig.~\ref{fig:SNIa_renormalization}, we adopt $\kappa_{\rm Ia}(\bar{\psi}_{\delta t}) = 0.8 + 0.2 \cdot \bar{\psi}_{\delta t}^{0.38}$, which is similar to Eq.~\ref{eq:error function} out to $\mathrm{log}_{10}(\bar{\psi}_{\delta t})\approx2$. The best-fit SFTs for the massive galaxies are much shorter and do not fit with the SPS constraints. This is shown by Fig~\ref{fig:SFT_different_k_SFR_relations} in the Appendix.

We note again that applying the same method of tuning the $\kappa_{\rm Ia}(\bar{\psi}_{\delta t})$ function under the assumption that the gwIMF is the invariant canonical IMF does not result in a meaningful solution. Assuming an invariant canonical IMF suggests shorter SFT values for more massive galaxies and/or longer SFTs for low-mass ellipticals, thus requiring a lower $\kappa_{\rm Ia}$ for massive galaxies and/or larger $\kappa_{\rm Ia}$ for low-mass galaxies to fit the SFT estimated by the SPS studies. This contradicts the expectation that massive galaxies with a more intense star formation activity should have a higher chance to form hard WD--WD and WD--stellar binary systems and SNIa events through the stellar-dynamical processing of binaries in massive star clusters which cannot form in galaxies with small SFRs.

\begin{figure}
    \centering
    \includegraphics[width=\hsize]{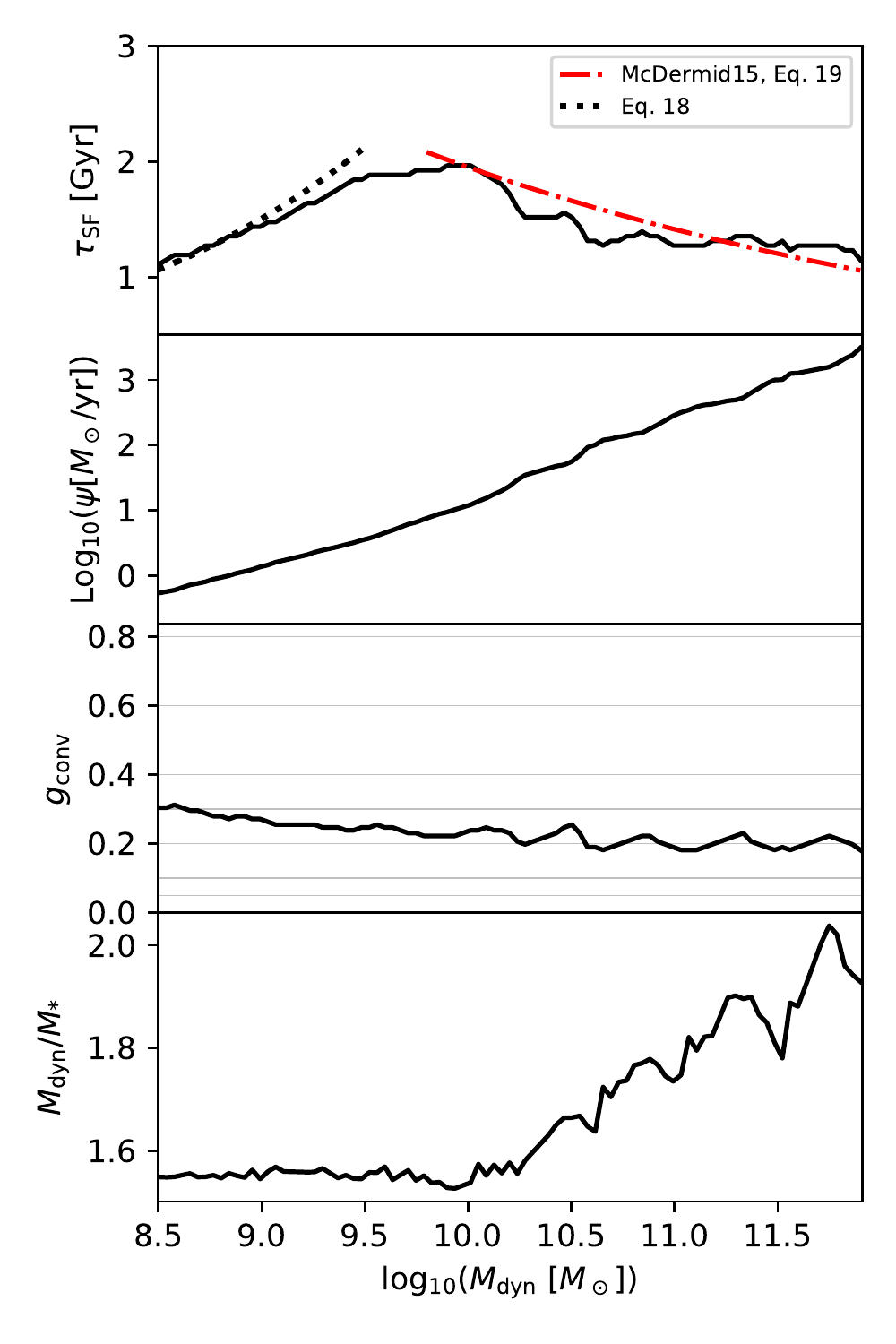}
    \caption{The SFT, SFR, gas-conversion parameter, and dynamical-mass-to-stellar mass ratio of our best-fit model, from the top to the bottom panel, for the error function model. The solid black line in the top panel is the same as the yellow ridgeline of Fig.~\ref{fig:best_fit_SFT_SNIa}. The black dotted and the red dash-dotted curves in the top panel follow the same relation as in Fig.~\ref{fig:best_fit_SFT_SNIa}, described by Eq.~\ref{eq:SFT fit} and \ref{eq:SFT McDermid}, respectively. The thin horizontal lines indicates $g_{\rm conv}$ values applied in our galaxy model (see Section~\ref{sec:Method}) where $g_{\rm conv}\approx1/3$ is about the star-formation efficiency found for embedded star clusters (Section~\ref{sec:IGIMF}).}
    \label{fig:best_fit}
\end{figure}

The SFT, SFR, gas-conversion parameter, and dynamical-to-stellar mass ratio of our best-fit models are shown in Fig.~\ref{fig:best_fit}. We find that roughly the same fraction of gas is consumed in our error function SNIa model (Eq.~\ref{eq:error function}) for galaxies with all different masses (the panel for $g_{\rm conv}$), being similar to the consumption fraction in embedded star clusters (Section~\ref{sec:IGIMF}), indicating a universal star-formation efficiency of about 1/3rd for monolithically (free-fall collapse) formed systems. We note that the wiggles are caused by the interpolation of the results from a limited number of galaxy models.
Other than the wiggles, the ratios of dynamical mass and stellar mass, $M_{\rm dyn}/M_*$, are higher for more massive galaxies because they have a more top-heavy gwIMF and more stellar remnants.
However, this result still needs to be translated to mass-to-light ratio, $M_{\rm dyn}/L_{\rm V}$, to compare with the observation. The difference between the $M_{\rm dyn}/L_{\rm V}$ values of massive and low-mass galaxies is even larger than the difference between the $M_{\rm dyn}/M_*$ values. This is because massive galaxies are also more metal-rich and subsequently have a more bottom-heavy gwIMF while the additional low-mass stars barely contribute to the V-band galaxy luminosity. A detailed discussion of this is given in Section~\ref{sec:Mass-to-light ratio}.

\section{Discussion}\label{sec:Discussions}

\subsection{Mass-to-light ratio}\label{sec:Mass-to-light ratio}

\begin{figure}
    \centering
    \includegraphics[width=\hsize]{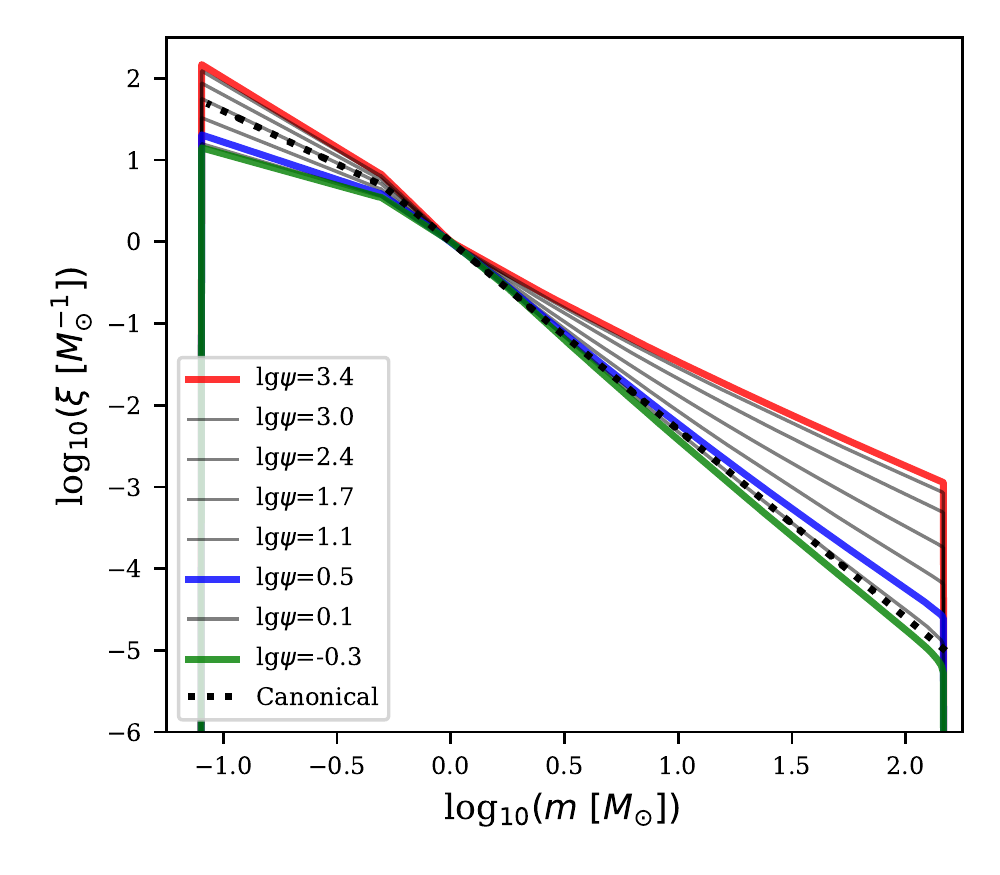}
    \caption{The gwIMFs, defined in Eq.~\ref{eq:IGIMF}, for the best-fit galaxies of our error function model (Eq.~\ref{eq:error function}) with masses $\mathrm{log}_{10}(M_{\rm dyn}/M_\odot)=8.5$, 9.0, 9.5, ..., 11.5, and 11.9 from the bottom to the top line. The galaxies with $\mathrm{log}_{10}(M_{\rm dyn}/M_\odot)=8.5$, 9.5, and 11.9 are highlighted by the green, blue, and red colour, respectively.
    The galactic SFRs ($\mathrm{lg}\psi \equiv \mathrm{log}_{10}\bar{\psi}_{\delta t}[M_\odot/\mathrm{yr}]$) of these models are given by the legend in units of $M_\odot$/yr. The red gwIMF has an approximated $\alpha_3$ of about 1.2 although the top-part of the IGIMF no longer preserves a power law form.}
    \label{fig:gwIMF_for_diff_SFR}
\end{figure}

\begin{figure}
    \centering
    \includegraphics[width=\hsize]{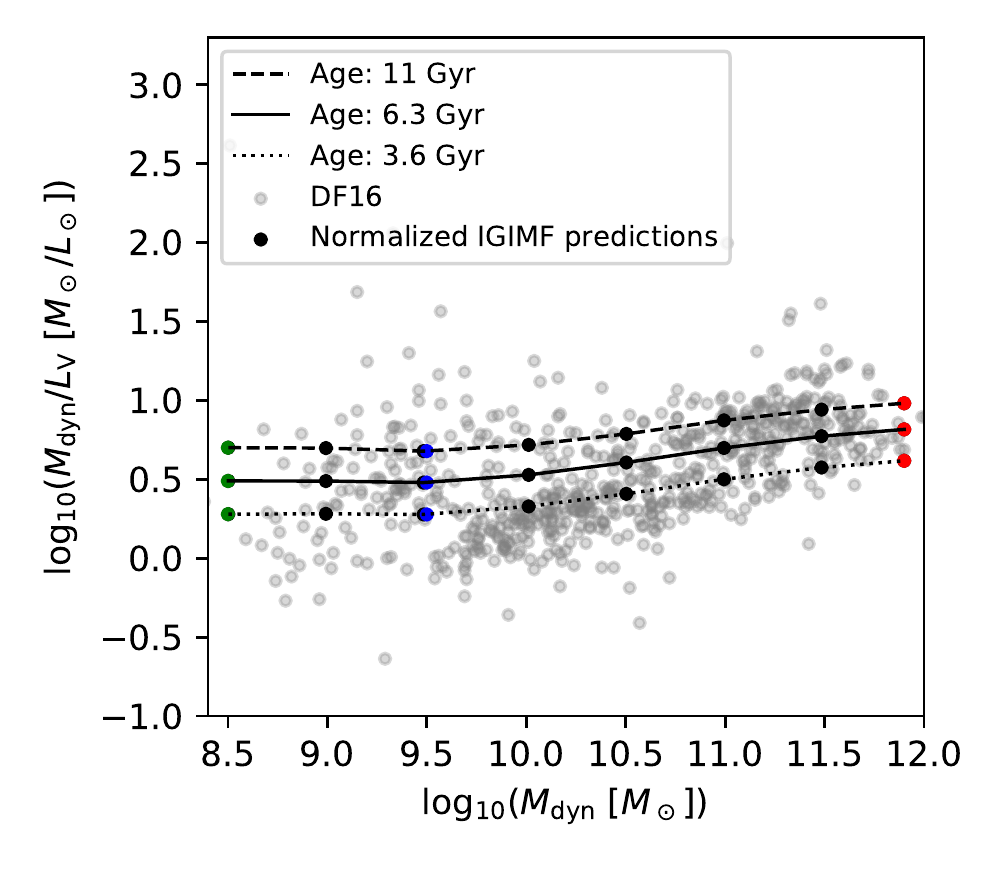}
    \caption{The normalized dynamical-mass-to-light ratio as a function of the dynamical mass of the galaxies (the grey circles are observational data given by \citealt{2016MNRAS.460.4492D}, DF16) at different galactic ages (denoted by the legend). The green, blue, and red points, i.e., galaxies with a final mass of $M_{\rm dyn}=10^{8.5}$, $10^{9.5}$, and $10^{11.9}~M_\odot$ are calculated with the gwIMF of the same colour given in Fig.~\ref{fig:gwIMF_for_diff_SFR}.}
    \label{fig:mass_to_light_ratio}
\end{figure}

We explore the V-band dynamical mass-to-light values, $M_{\rm dyn}/L_{\rm V}$, of the best-fit models described in Section~\ref{sec:SFT with variable SNIa rate}. 
To simplify the calculation, we apply a simple-population approximation which assumes that all the stars in a galaxy have the same age and metallicity, where the gwIMF of the galaxy is calculated with the best-fit SFR (second panel of Fig.~\ref{fig:best_fit}) and the observed mean stellar metallicity (upper panel of Fig.~\ref{fig:data}). The simple-population gwIMFs of a few galaxies with different masses are shown in Fig~\ref{fig:gwIMF_for_diff_SFR}.
The $M_{\rm dyn}/L_{\rm V}$ value of a galaxy can be calculated for these gwIMFs for a given age. We apply $t=3.6$, $6.3$, and $11$ Gyr corresponding to the mean and standard deviation of the estimated galaxy ages of the observed galaxy population (see Section~\ref{sec:Observational constrain}).
Since in the GalIMF code, the luminosity is calculated using the main-sequence stellar luminosities only, as described by \citet[their eq. 1]{2019A&A...629A..93Y}, the real galactic luminosity 
would be systematically higher due to the AGB stars such that
the $M_{\rm dyn}/L_{\rm V}$ ratios would be lower than the calculation described here. Therefore, we divide the $M_{\rm dyn}/L_{\rm V}$ ratios by the same factor of 1.425 for all galaxies to normalize the $M_{\rm dyn}/L_{\rm V}$ values such that the galaxy with a mass of $M_{\rm dyn}=10^{9.5} M_\odot$ has $M_{\rm dyn}/L_{\rm V}=3$ in Solar units. The E-MILES stellar population model \citep{2016MNRAS.463.3409V} combined with the GalIMF code will be applied to give more accurate $M_{\rm dyn}/L_{\rm V}$ values in the near future.
The results are shown and compared with the mass-to-light ratios from direct observations in Fig.~\ref{fig:mass_to_light_ratio}. 
The agreement is remarkable: the here computed best-fit models follow the data trend very well with the implication that the masses of the most-massive E galaxies are significantly dominated by faint M~dwarfs as well as with stellar remnants in agreement with the spectroscopic studies (see discussion in \citealt{2021MNRAS.500.4153L} and references therein).
These massive galaxies have higher $M_{\rm dyn}/L_{\rm V}$ values due to the fact that their bulk super-solar metallicity leads to a bulk bottom-heavy gwIMF as is shown by the red line in Fig~\ref{fig:gwIMF_for_diff_SFR}.

\begin{figure}
    \centering
    \includegraphics[width=\hsize]{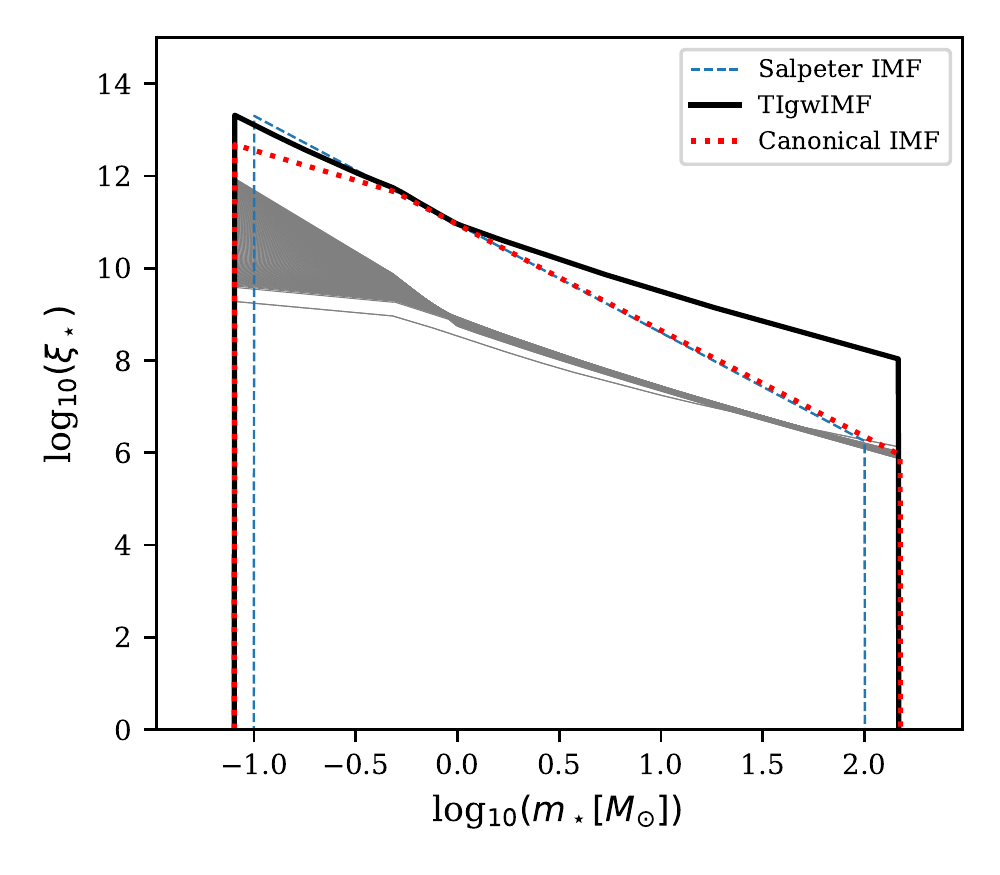}
    \caption{The gwIMF, defined in Eq.~\ref{eq:IGIMF}, of each 10 Myr star formation epoch (thin lines) and the time-integrated gwIMF for all formation epochs, TIgwIMF, for a galaxy with a final mass of $M_{\rm dyn}=10^{11.9} M_\odot$. This TIgwIMF corresponds to the red model in Fig.~\ref{fig:mass_to_light_ratio}. The corresponding approximating simple-population IGIMF model is shown in Fig.~\ref{fig:gwIMF_for_diff_SFR} as a red line. The gwIMF for the first and the last star formation epoch are highlighted by slightly thicker lines. The canonical IMF as given by \citet{2001MNRAS.322..231K} and the power-law IMF given by \citet{1955ApJ...121..161S} are shown as the red-dotted and blue-dashed lines, respectively.}
    \label{fig:galaxy_evolution_fig_TIgwIMF_82}
\end{figure}

Unlike the gwIMF calculated here using the simple-population approximation and the observed mean stellar metallicity, the real gwIMF evolves through each star formation epoch. At the onset of the formation of all galaxies, the gwIMFs were all similarly bottom-light with a top-heaviness which depends on the SFR with the gwIMF becoming bottom-heavier as the metallicity of the galaxy increases. For example, Fig.~\ref{fig:galaxy_evolution_fig_TIgwIMF_82} plots the gwIMF for each 10 Myr star formation epoch and also for the integrated gwIMF of all star formation epochs, i.e., a time-integrated gwIMF (TIgwIMF) for a galaxy with $\mathrm{log}_{10}(\bar{\psi}_{\delta t}[M_\odot/\mathrm{yr}])=3.4468$, $\tau_{\rm SF}=1.19$~Gyr, $g_{\rm conv}=0.1898$, $[Z/X](t=0)=-6$, and a final mass of $10^{11.9}~M_\odot$, corresponding to the high-SFR red gwIMF in Fig.~\ref{fig:gwIMF_for_diff_SFR}. The TIgwIMF for an evolved stellar population with different ages reproduces the gwIMF calculated using the observed mean stellar metallicity well, indicating that it is valid to calculate $M_{\rm dyn}/L_{\rm V}$ with the simple-population approximation.
Similarly, the TIgwIMFs for galaxies with a mass of $10^{8.5}$ or $10^{9.5}$ $M_\odot$ are shown in Fig.~\ref{fig:galaxy_evolution_fig_TIgwIMF_127} and \ref{fig:galaxy_evolution_fig_TIgwIMF_217}, respectively, in the Appendix. Their TIgwIMFs are similar to the canonical IMF, leading to a standard $M_{\rm dyn}/L_{\rm V}$ value. We refer the reader to  \cite{2016MNRAS.463.1865D} and \cite{2019MNRAS.490..848D} for further analysis of the dynamical mass to light ratios of ETGs within the IGIMF framework.

\subsection{Stellar yield tables}\label{sec:Stellar yield}

The estimated SFT for a galaxy with a given mass depends sensitively on the adopted stellar yield table. But since the yield is applied to all galaxies uniformly, the general trend (the slope) of the $\tau_{\rm SF}$--$M_{\rm dyn}$ relation and the scientific conclusions remain the same. We have tested that the effect of applying a different yield table on the resulting $\tau_{\rm SF}$--$M_{\rm dyn}$ relation is similar to having a uniform bias of the observed galactic metal abundance. This has been demonstrated in our previous work \citep[their section 5.1]{2019A&A...632A.110Y}.

\subsection{The application of closed-box models}\label{sec:Closed-box model}

The closed-box modelling approach has the advantage that a large range of parameters (e.g. the galaxy mass, formation time-scale, gwIMF properties) is computationally reachable, while cosmological hydrodynamical galaxy-formation models with star formation and feedback processes are computationally too costly to allow a comprehensive investigation of parameter-space.

The observations of E~galaxies have shown that in-situ star formation contributes a large fraction of the stellar population while galaxy mergers are invoked to gradually increase the size and the mass of the most massive E~galaxies over a long period of time, building up the outer regions of the galaxies
(\citealt{2006ARA&A..44..141R,2010ApJ...725.2312O,2013MNRAS.436.3507N,2015MNRAS.448.3484M,2015ApJ...808...79F,2018MNRAS.475.3700M,2019MNRAS.487.4939M,2020MNRAS.491.3562Z,2020A&A...644A.117L}, for a discussion see section~8.2 in \citealt{2020MNRAS.498.5652K}). Since the timescale for most of the stars to form is short, the amount of galactic gas inflow and/or outflow cannot be significant to strongly affect the abundances of these stars (section~5.1 in  \citealt{2019A&A...629A..93Y}). For this reason, the central region of giant ellipticals can be reasonably approximated by the closed-box model with neither galactic in- nor outflow \citep{2012ceg..book.....M}. 

On the other hand, the theory of standard-dark-matter-based cosmology predicts that more massive galaxies form from more mergers instead of the in-situ star formation described by the closed-box model (e.g. \citealt{2019arXiv191009552M}), in conflict with the observational studies.
Reproducing the abundance scaling relations within the hierarchical galaxy formation framework has long been a problem. By tuning the star formation and feedback parameters, it has been shown possible for the simulated galaxies to broadly fulfil the observed mass--metallicity relation for galaxies at different redshifts and also the radial metallicity gradient of resolved galaxies \citep{2016MNRAS.461.1760H,2018MNRAS.477L..16T,2019MNRAS.484.5587T,2020arXiv200710993H}. But the reproduction for the most massive galaxies is not entirely satisfactory. The observed $M_{\rm dyn}$--stellar-metallicity relation has a higher metallicity and a smaller scatter of the metallicity value for the more massive galaxies, which is not seen in the simulations (e.g. the comparison between observed and predicted relation in \citealt[their fig. 7]{2010MNRAS.402..173A} and \citealt[top panel of their fig. 11]{2018MNRAS.479.5448B}). In addition, the $\alpha$-enhancement abundances of the massive galaxies are not simultaneously reproduced with the metal-enhancement \citep{2017MNRAS.464.4866O,2017MNRAS.466L..88D,2019A&A...629L...3R}. This turns out to be a tricky problem requiring non-canonical solutions including a non-canonical IMF (\citealt{2005MNRAS.358.1247N,2009MNRAS.400.1347C,2010MNRAS.402..173A,2015MNRAS.446.3820G,2017MNRAS.464.3812F,2018MNRAS.479.5448B,2019MNRAS.482..118G}), non-canonical SNIa incidence (\citealt{2010MNRAS.402..173A}), non-canonical stellar yields (see Section~\ref{sec: alternative solutions}), or differential galaxy winds (\citealt{2013MNRAS.435.3500Y}).
Hence, the hierarchical formation scenario appears to need further tuning and parameter-addition of the baryonic physics parameters.

Even if the merger scenario for more massive galaxies is correct, it does not necessarily disqualify the closed-box galaxy chemical evolution models. The satellite galaxies that merged into and contributed a large mass fraction to the massive galaxies should not have the same property as the low-mass galaxies surviving to date and have undergone processing over a Hubble time. Otherwise, the massive galaxies would not preserve their unique higher metallicity. The merged satellite galaxies are thought to have been more metal-rich than the central galaxies of the same mass \citep{2010MNRAS.407..937P} since they have lived in the vicinity of a massive galaxy or in a group of galaxies with a higher total mass and are thus not isolated but subject to the chemical enrichment of the entire gravitationally bound gas cloud and/or galaxy group \citep{2017MNRAS.464..508B} that later merged into a single massive galaxy where the enriched galactic outflows of each galaxy in the group are recycled back within the group. In this sense, the closed-box approximation can broadly describe the chemical evolution of the massive E~galaxies when considering the entire progenitor galaxy group as a whole, while the exact dynamical evolution and assembly history are not a concern here. Certainly, this description is only an approximation and not all galaxies can be well-represented by the closed-box model. There have been observations of outliers with abnormal abundance ratios unexpected by the closed-box model (e.g. the iron-poor population documented in \citealt{2015MNRAS.448.3484M} and \citealt{2020ApJ...897L..42J}). Such galaxies can only be explained by a more complicated formation scenario, for example, through a larger amount of primordial gas inflow or other mechanisms that modify the chemical abundance. Nevertheless, taken at face value, the properties of ETGs are simplest and most concisely understood through the monolithic post-Big-Bang gas-cloud collapse theory for their origin \citep{2020MNRAS.498.5652K}.

\subsection{Alternative solutions assuming a canonical IMF}\label{sec: alternative solutions}

Can different gwIMF and/or different stellar yields, different SNIa DTDs, gas mixing and expulsion physics, or a combination of these reproduce the observed chemical abundance ratios of E~galaxies as is introduced in Section~\ref{sec:intro}? Thus, is it possible that the most massive E~galaxies reach their high mean stellar [Mg/Fe] value with a SFT that is no shorter than about 1 Gyr assuming that the gwIMF is invariant and canonical?

Let us first consider the possibility of a variable $\kappa_{\rm Ia}$ as a function of galaxy mass or the SFR when the IMF is canonical and invariant:
In order to reproduce the high [Mg/Fe] ratios of massive E~galaxies, smaller $\kappa_{\rm Ia}$ values would be required and, therefore, contradict the observation. For example, \citet{2010MNRAS.402..173A} find, with the canonical and a slightly top-heavy gwIMF, that the [$\alpha$/Fe] values of massive galaxies require too low a SNIa production efficiency which is hard to reconcile with the observed SNIa rate of the star-forming galaxies. Also, the expectation that high-density metal-rich star-forming regions in a massive E~galaxy are likely to have a higher fraction of short-period binary stars and thus higher $\kappa_{\rm Ia}$ values (\citealt{2002ApJ...571..830S}, \citealt{2005A&A...441.1055G} their section 2 and references therein, and \citealt{2018MNRAS.479.3563F}). 
Furthermore, if the bulk stellar population of E~galaxies forms through a monolithic collapse of a post-Big-Bang gas cloud, as is suggested by much observational evidence and theoretical modelling (\citealt{2020MNRAS.498.5652K} and references therein), then these galaxies would be forming extremely massive star-burst clusters in-line with the observed extragalactic correlation between the SFR and cluster masses \citep{2004MNRAS.350.1503W,2013ApJ...775L..38R,2017A&A...607A.126Y}. As massive star-burst clusters are SNIa factories \citep{2002ApJ...571..830S} and given the above arguments by \cite{2005A&A...441.1055G}, it appears likely that $\kappa_{\rm Ia}$ indeed increases with increasing E~galaxy mass. We show in Section~\ref{sec:SFT with variable SNIa rate} that $\kappa_{\rm Ia}$ indeed needs to increase for the more massive E~galaxies under the framework of the IGIMF theory.

The other solutions include a more efficient loss of the elements produced by SNIa than those produced by type~II supernovae. This could help to increase the galactic [Mg/Fe] values, allowing for longer SFT values. However, such a supernova-type-dependent wind would be more prominent in dwarf galaxies which have shallower potential wells \citep{2001MNRAS.322..800R} than for the massive galaxies. Given the large E~galaxy masses, here we apply the closed-box model and neglect any galactic winds. Also, the stellar yields may change if the average stellar rotation speed differs in different galaxies due to, for example, a difference in the stellar metallicity \citep{2019MNRAS.490.2838R,2020arXiv200505717R}. This possibility requires further investigation.



\subsection{Comparison with the SFTs estimated with other methods}\label{sec:Caveat on SPS}

The downsizing behaviour of galaxy formation is supported independently by dynamical simulations (\citealt{2009ApJ...705..650W,2021arXiv210112226J,Eappen+2021}), chemical evolution models (\citealt{2005ApJ...621..673T,2010MNRAS.404.1775T}), and SPS studies (\citealt{2011MNRAS.418L..74D,2015MNRAS.448.3484M}). However, it is not trivial to confirm or falsify the here estimated $\tau_{\rm SF}$--$M_{\rm dyn}$ relation (the yellow ridgeline in Fig.~\ref{fig:best_fit_SFT_SNIa}) using different methods since no observation of the SFH is direct and there is a large systematic uncertainty related to all the methods.

For example, the chemical evolution models in this study assume a perfect galaxy with no gas flows, a constant SFR during the star formation epoch, and zero SFR before and after that. These approximations are likely to be valid only for massive E~galaxies. The observation of dwarf galaxies with a low total stellar luminosity can be significantly dominated by their most recent star formation from a gas reservoir polluted by the local environment, invalidating our model assumptions. In addition, as mentioned in Section.~\ref{sec:SFT with variable SNIa rate}, the real fluctuating SFR in dwarf galaxies rather than the assumed smooth SFH would lead to a more top-heavy gwIMF, and therefore to a higher estimated SFT than our results shown in Fig.~\ref{fig:best_fit_SFT_SNIa}.

The SFTs suggested by the SPS studies, such as by \citet{2015MNRAS.448.3484M}, are subject to the assumed stellar population model, dust attenuation law, and ad hoc preference on a smooth SFH. The assumed gwIMF and stellar abundance ratios of different stellar populations being synthesised are prior assumptions that introduce biases to the resulting SFH. Due to the degenerate nature of inferring the SFH from the integrated galaxy light, the solution is not unique \citep{2014MNRAS.444.3408Y,2020A&A...635A.136A}. The SPS method is not able to detect strong short-term fluctuations and the resulting SFH is strongly affected by the prior assumed SFH shape and/or smoothness constraints \citep{2015MNRAS.448.3484M,2017ApJ...838..127I,2018MNRAS.480.4379C,2019ApJ...873...44C,2019ApJ...876....3L,2019ApJ...879..116I,2020arXiv200603599L,2021arXiv210212494T}. With higher SFRs extending for shorter times, the total stellar mass formed with the same age and metallicity is preserved while the gwIMF can be very different if it depends on the star formation intensity as is suggested by the IGIMF theory (Section~\ref{sec:IGIMF}). As mentioned in \citet{2019A&A...632A.110Y}, with non-unique SPS solutions, there is no guarantee that the preferred solution (imposed by the prior) reproduces the observed abundance in a self-consistent chemical evolution simulation (see \citealt{2020MNRAS.498.5581B} as a first step to consider the SPS and the metal-evolution of the galaxy consistently). 

The SFH measurements would be much more reliable and encouraging if the estimations from all different methods would be consistent with each other. More recent studies of high-redshift galaxies by \citet{2019MNRAS.484.2281S} broadly agree with the downsizing relation suggested by \citet{2015MNRAS.448.3484M}. Independent methods such as measuring the SFH with pixel colour-magnitude diagrams (CMD, \citealt{2019ApJ...876...78C}) could also be helpful supplementary evidence supporting the downsizing SFH of massive ETGs. The uncertainty of such a method is large, but better-resolved pixel CMD images have the potential to reveal the SFH fluctuations that are hidden in integrated-light observations. Larger future observational platforms are likely to improve our understanding of the assembly of ETGs.

With the consensus that the SFT of massive ETGs should be longer than about 1~Gyr, we are able to robustly exclude the canonical invariant IMF model in \citet{2019A&A...632A.110Y}. We demonstrate in this work that an agreement between the estimate of the SFTs of massive ETGs using the SPS method on the one hand side and chemical evolution studies on the other can be achieved if the IGIMF theory is applied (Fig.~\ref{fig:best_fit_SFT_SNIa}). The required higher SNIa production efficiency for massive ETGs (Fig.~\ref{fig:SNIa_variation_IGIMF_varKappa}) agrees beautifully with the independent theoretical expectation \citep{2002ApJ...571..830S} and observation \citep{2021MNRAS.502.5882F} for the first time. On the other hand, the SFHs of dwarf galaxies which have much shallower potentials, non-canonical gwIMFs, and dust is harder to determine. We avoid fine-tuning our model or to introduce additional model parameters. At face value, our study suggests a short SFT for the typical early-type dwarf galaxies.

\section{Conclusion}\label{sec:Conclusion}

In this work, we adopt for the first time an environment-dependent IMF, that is, the IGIMF theory, in our galaxy chemical evolution model, GalIMF, to estimate the formation timescale of E~galaxies using both stellar mean metallicity and [Mg/Fe] constrains. The IMF changes as a consequence of the changing metallicity and SFR of the galaxy, which in turn affects the chemical enrichment. In Eq.~\ref{eq:N_Ia}, we formulate how the total number of SNIa per unit mass of stars formed, $N_{\rm Ia}$, should vary with the IMF shape. The results assuming an invariant overall SNIa realisation parameter, i.e., our fiducial SNIa model, are shown in Fig.~\ref{fig:SNIa_variation_IGIMF}, accounting for the observed large scatter of the observational estimates. A higher $N_{\rm Ia}$ is inferred for metal-poor galaxies that have a bottom-light IMF. While a lower $N_{\rm Ia}$ is inferred for low-SFR galaxies that have a top-light gwIMF.

The environment-dependent gwIMF affects the SFT evaluation significantly. The resulting SFT is longer for massive E~galaxies than the value estimated by the SPS studies. A natural explanation of this discrepancy is that the real total number of SNIa per stellar mass formed is higher for more massive galaxies than the number estimated by the fiducial SNIa model because they had more intense and denser star-forming activities. A plausible 2.5 times increase of the overall SNIa realisation parameter for the most massive galaxies (the error function model shown in Fig.~\ref{fig:SNIa_renormalization}) as suggested by \citet{2018MNRAS.479.3563F} and \citet{2021MNRAS.502.5882F} can resolve this discrepancy and fits the SFT values obtained from SPS. This is consistent with massive galaxies having formed more massive star-burst clusters as a consequence of their higher SFRs because such clusters are factories for SNIa progenitors and are also relevant for the rapid emergence of super-massive black holes \citep{2002ApJ...571..830S,2020MNRAS.498.5652K}. Future dynamical studies may be able to further test this prediction with N-body simulations. On the other hand, adopting the invariant canonical IMF in the galaxy chemical evolution model would not be able to reproduce these galaxy SFTs and metal abundance constraints because such a model would require a lower SNIa production efficiency for massive galaxies than for low-mass galaxies.

The IGIMF theory with a more bottom-light gwIMF for the low-SFR galaxy also suggests that the SFT of E~galaxies should decrease for lower mass galaxies. Considering that the SFT also decreases for the most massive galaxies as is suggested by the SPS studies, galaxies with a mass of about $10^{9.5} M_\odot$ may have the longest SFT (Fig.~\ref{fig:best_fit}). This result implies that most stars in more massive E~galaxies were formed in a shorter post-Big-Bang gas-cloud collapse timescale, with roughly the same fraction (1/3rd) of gas converted to stars and a higher mass-to-light ratio, while lower-mass galaxies tend to lose their gas supply and quench star formation due to feedback in a shorter timescale. Examples of this may be the ultra-diffuse galaxy Dragonfly~44 \citep{2019ApJ...884L..25H} and ultra-faint dwarf galaxy Bo\"otes~I \citep{2020A&A...637A..68Y}.

We conclude that the IGIMF theory along with an increased production efficiency of SNIa for more massive E~galaxies (described by Eq.~\ref{eq:N_Ia}, \ref{eq:SNIa_renormalization}, and \ref{eq:error function}) is able to explain the observed stellar populations of E~galaxies successfully, thereby also self-consistently accounting for the large dynamical mass-to-light ratios (Fig.~\ref{fig:mass_to_light_ratio}) and the bottom-heavy stellar mass function in massive E~galaxies. A noteworthy aspect of the theory of elliptical galaxy formation and evolution arrived at here is that many observed properties come together naturally and are in fact straightforwardly explained: the higher incidence of SNIa events in galaxies that formed over the downsizing time with high SFRs, the downsizing-implied systematic galaxy-wide IMF evolution accounting for the metallicity and $\alpha$-element enhancement and the implied rapid formation of supermassive black holes \citep{2020MNRAS.498.5652K}, all are a consequence of the application of the IGIMF theory to the monolithic collapse of post-Big-Bang gas clouds \citep{Eappen+2021}.


\begin{acknowledgements}
ZY acknowledges financial support from the China Scholarship Council (CSC, file number 201708080069). TJ acknowledges support through the European Space Agency fellowship programme. PK acknowledges support from the Grant Agency of the Czech Republic under grant number 20-21855S. The development of our chemical evolution model applied in this work benefited from the International Space Science Institute (ISSI/ISSI-BJ) in Bern and Beijing, thanks to the funding of the team “Chemical abundances in the ISM: the litmus test of stellar IMF variations in galaxies across cosmic time” (Donatella Romano and Zhi-Yu Zhang).
\end{acknowledgements}

\bibliographystyle{aa} 
\bibliography{references}

\begin{appendix}\label{sec:appendix}

\section{SNIa production efficiency calculated with different IGIMF formulations and SNIa progenitor mass range}\label{sec:supplementary plots}

\begin{figure}[H]
    \centering
    \includegraphics[width=\hsize]{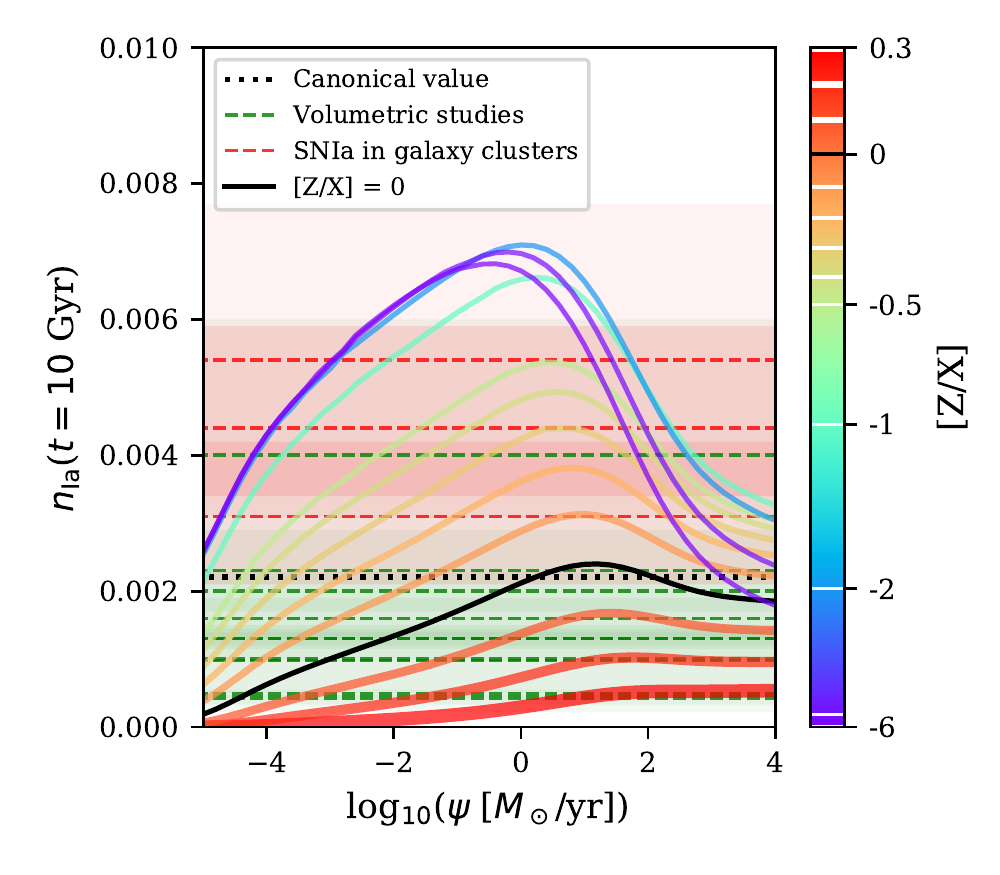}
    \caption{The 10 Gyr time-integrated number of SNIa per unit stellar mass formed. Same as Fig.~\ref{fig:SNIa_variation_IGIMF} but with a SNIa progenitor mass range from 2 to 8 $M_\odot$ instead of 3 to 8 $M_\odot$.}
    \label{fig:SNIa_variation_IGIMF_2t8}
\end{figure}

\begin{figure}[H]
    \centering
    \includegraphics[width=\hsize]{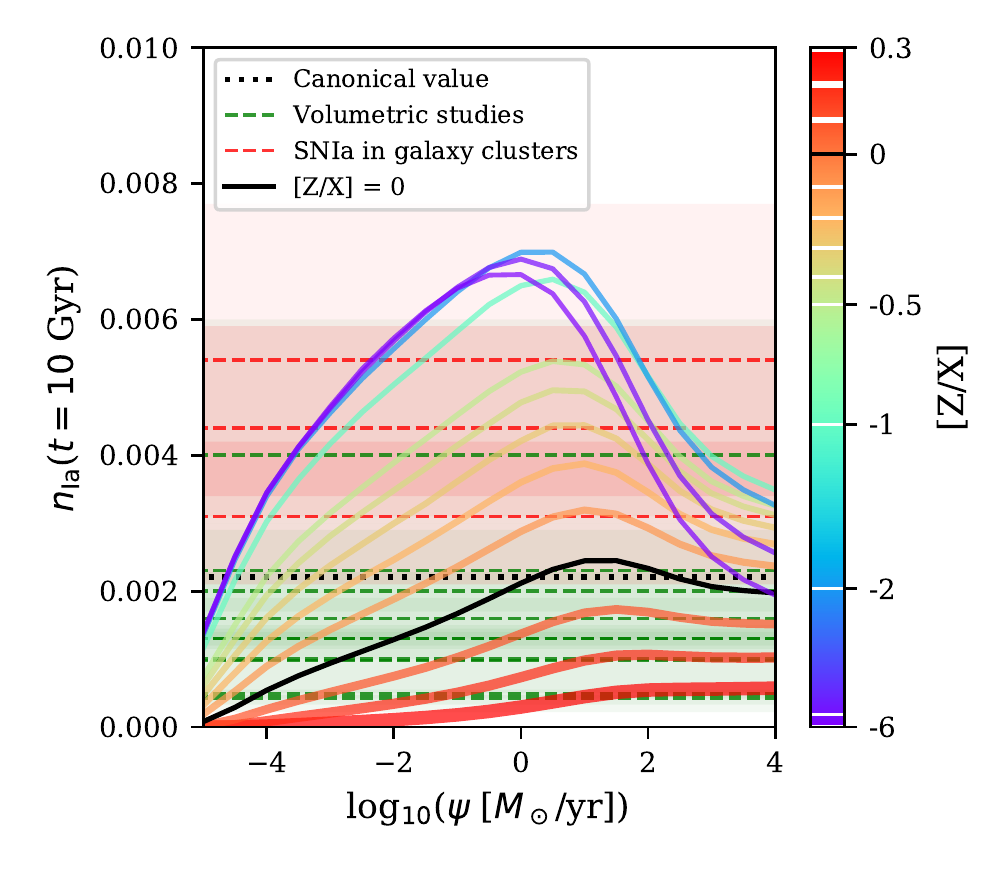}
    \caption{The 10 Gyr time-integrated number of SNIa per unit stellar mass formed. Same as Fig.~\ref{fig:SNIa_variation_IGIMF} but with a SNIa progenitor mass range from 3 to 8 $M_\odot$ for the primary star and 1.5 to 8 $M_\odot$ for the secondary star instead of 3 to 8 $M_\odot$ for both stars.}
    \label{fig:SNIa_variation_IGIMF_3a15t8}
\end{figure}

The SNIa production efficiency calculated with a different SNIa progenitor mass range from 2 to 8 $M_\odot$ is shown in Fig.~\ref{fig:SNIa_variation_IGIMF_2t8}. The efficiency is higher for the low-mass metal-poor galaxies compared to Fig.~\ref{fig:SNIa_variation_IGIMF}. Results are similar if the considered mass ranges for the primary and the secondary star are different, e.g., from 3 to 8 $M_\odot$ and 1.5 to 8 $M_\odot$, respectively (Fig.~\ref{fig:SNIa_variation_IGIMF_3a15t8}).

\begin{figure}[H]
    \centering
    \includegraphics[width=\hsize]{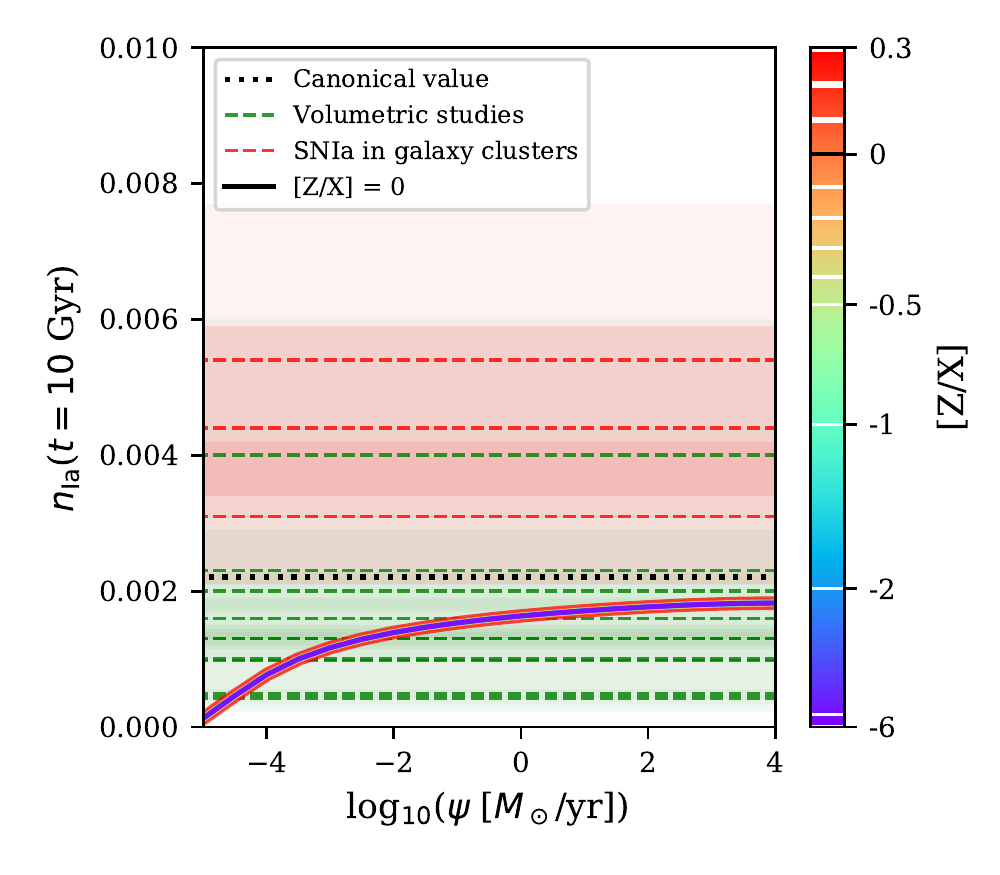}
    \caption{The 10 Gyr time-integrated number of SNIa per unit stellar mass formed. Same as Fig.~\ref{fig:SNIa_variation_IGIMF} but adopt the IGIMF1 model of \citet{2018A&A...620A..39J} instead of Eq.~\ref{eq:alpha18} and \ref{eq:alpha3}.}
    \label{fig:SNIa_variation_IGIMF1}
\end{figure}

\begin{figure}[H]
    \centering
    \includegraphics[width=\hsize]{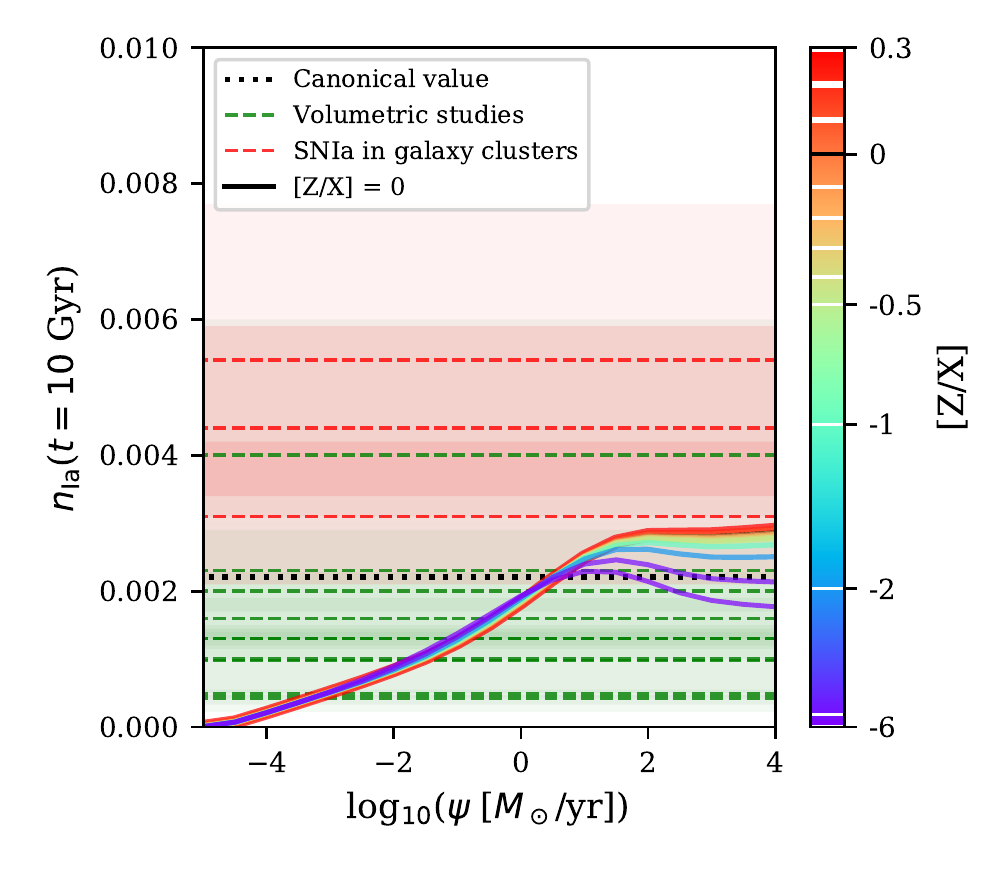}
    \caption{The 10 Gyr time-integrated number of SNIa per unit stellar mass formed. Same as Fig.~\ref{fig:SNIa_variation_IGIMF} but adopt the IGIMF2 model of \citet{2018A&A...620A..39J} instead of Eq.~\ref{eq:alpha18} and \ref{eq:alpha3}.}
    \label{fig:SNIa_variation_IGIMF2}
\end{figure}

\begin{figure}[H]
    \centering
    \includegraphics[width=\hsize]{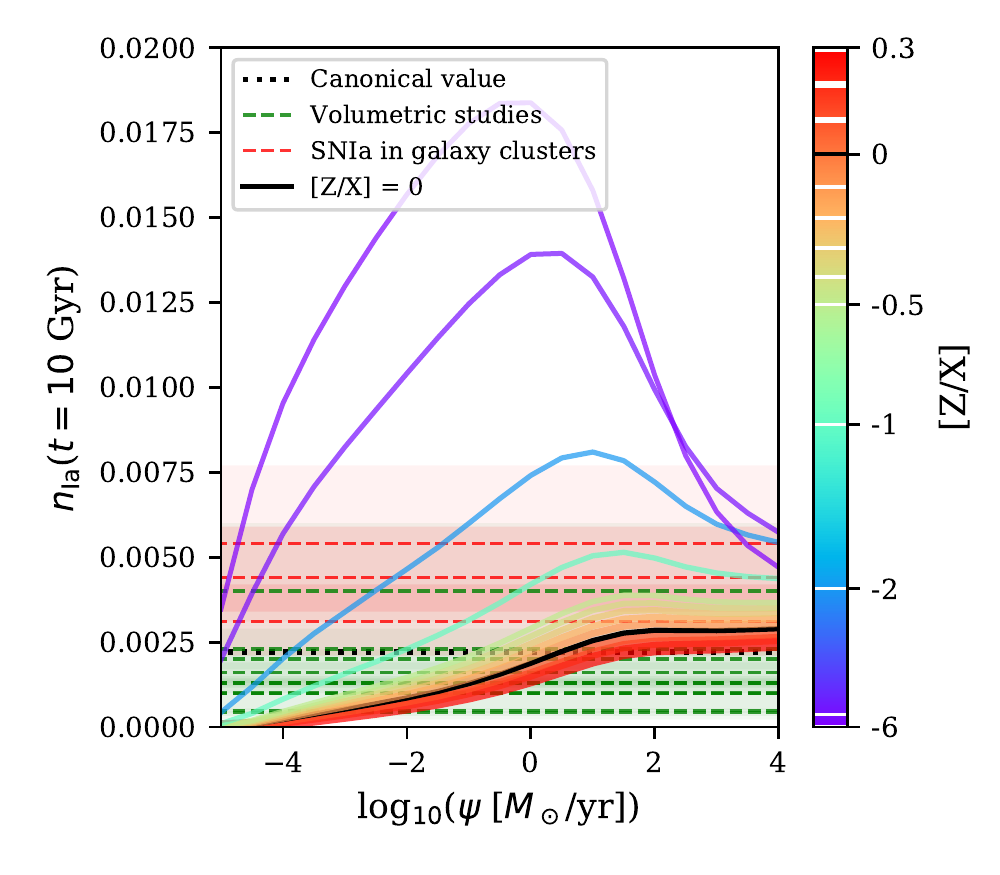}
    \caption{The 10 Gyr time-integrated number of SNIa per unit stellar mass formed. Same as Fig.~\ref{fig:SNIa_variation_IGIMF} but adopt the IGIMF3 model of \citet{2018A&A...620A..39J} instead of Eq.~\ref{eq:alpha18} and \ref{eq:alpha3}.}
    \label{fig:SNIa_variation_IGIMF3}
\end{figure}

Fig.~\ref{fig:SNIa_variation_IGIMF1} to \ref{fig:SNIa_variation_IGIMF3} give the SNIa production efficiency calculated for the IGIMF1, IGIMF2, and IGIMF3 formulations as summarised in \citet[their table 3]{2018A&A...620A..39J}. Models with different metallicity are overlapping in Fig.~\ref{fig:SNIa_variation_IGIMF1} because the IGIMF1 model is independent of metallicity.
The IGIMF2 model has an invariant IMF for low-mass stars. The IGIMF3 model has an IMF dependence on $[Z]$ instead of $Z$ in Eq.~\ref{eq:alpha18} for low-mass stars. 

The results shown in this section following Eq.~\ref{eq:N_Ia} can be used to exclude theories of how the IMF varies.



\section{The power-law function model}
\begin{figure}[H]
    \centering
    \includegraphics[width=\hsize]{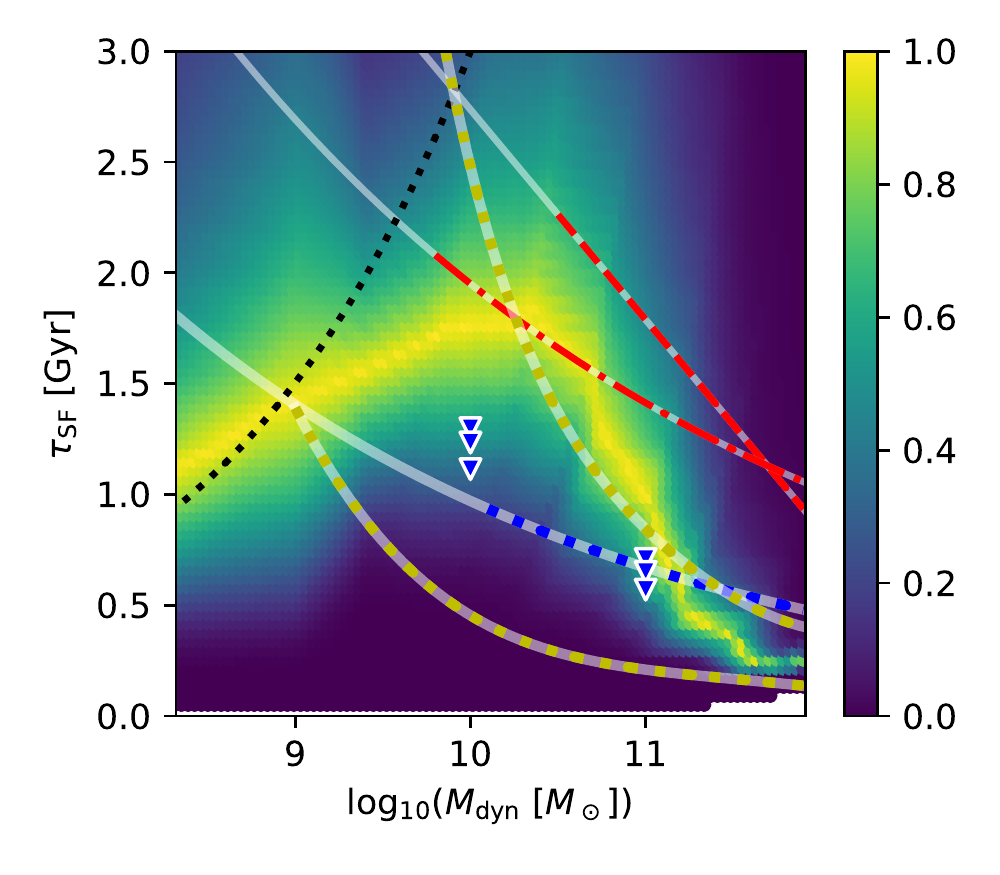}
    \caption{$\tau_{\rm SF}$ as a function of $M_{\rm dyn}$ for the power-law function model. Same as Fig.~\ref{fig:best_fit_SFT} but the results shown by the colour map adopt the power-law $\kappa_{\rm Ia}$--$\bar{\psi}_{\delta t}$ relation shown in Fig.~\ref{fig:SNIa_renormalization}. 
    }
    \label{fig:SFT_different_k_SFR_relations}
\end{figure}

Fig.~\ref{fig:SFT_different_k_SFR_relations} shows the best-fit $\tau_{\rm SF}$ as a function of $M_{\rm dyn}$ for the power-law function model (see Fig.~\ref{fig:SNIa_renormalization}) by the yellow ridgeline.

\section{The evolution of the gwIMFs}

\begin{figure}[H]
    \centering
    \includegraphics[width=\hsize]{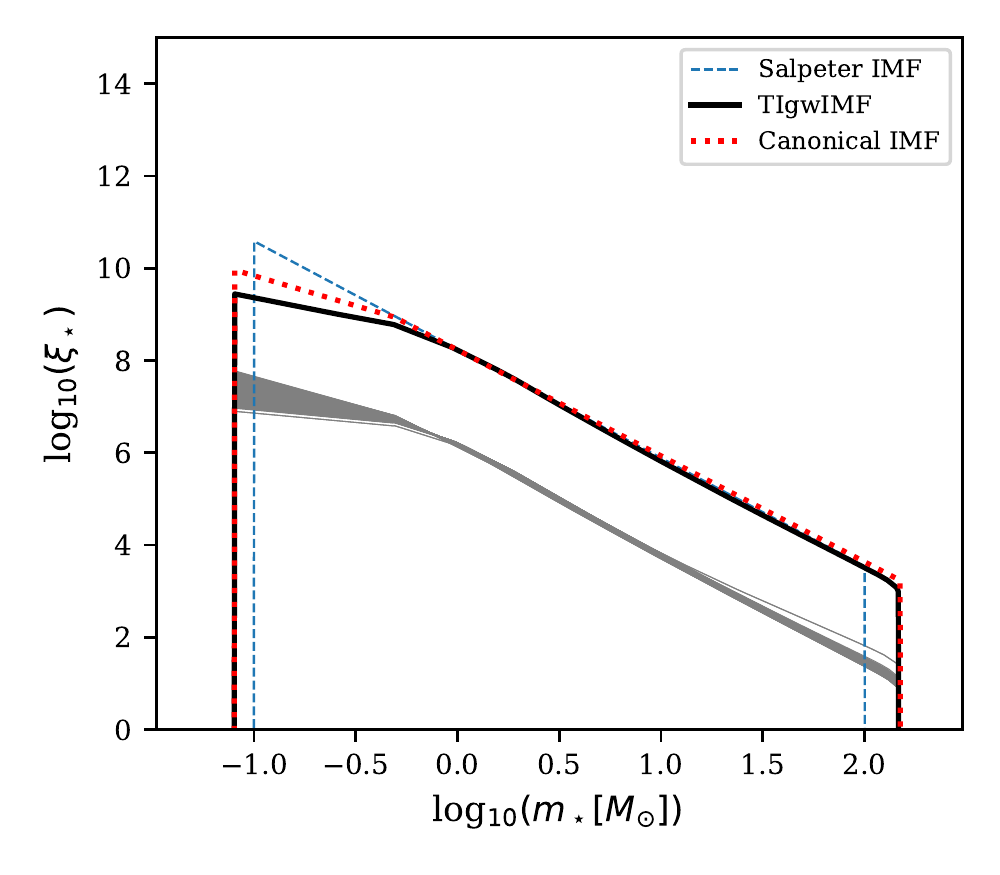}
    \caption{Same as Fig.~\ref{fig:galaxy_evolution_fig_TIgwIMF_82} but for a galaxy with a final mass of $M_{\rm dyn}\approx 10^{8.5} M_\odot$. This TIgwIMF is well represented by the green model in Fig.~\ref{fig:gwIMF_for_diff_SFR}.}
    \label{fig:galaxy_evolution_fig_TIgwIMF_127}
\end{figure}

\begin{figure}[H]
    \centering
    \includegraphics[width=\hsize]{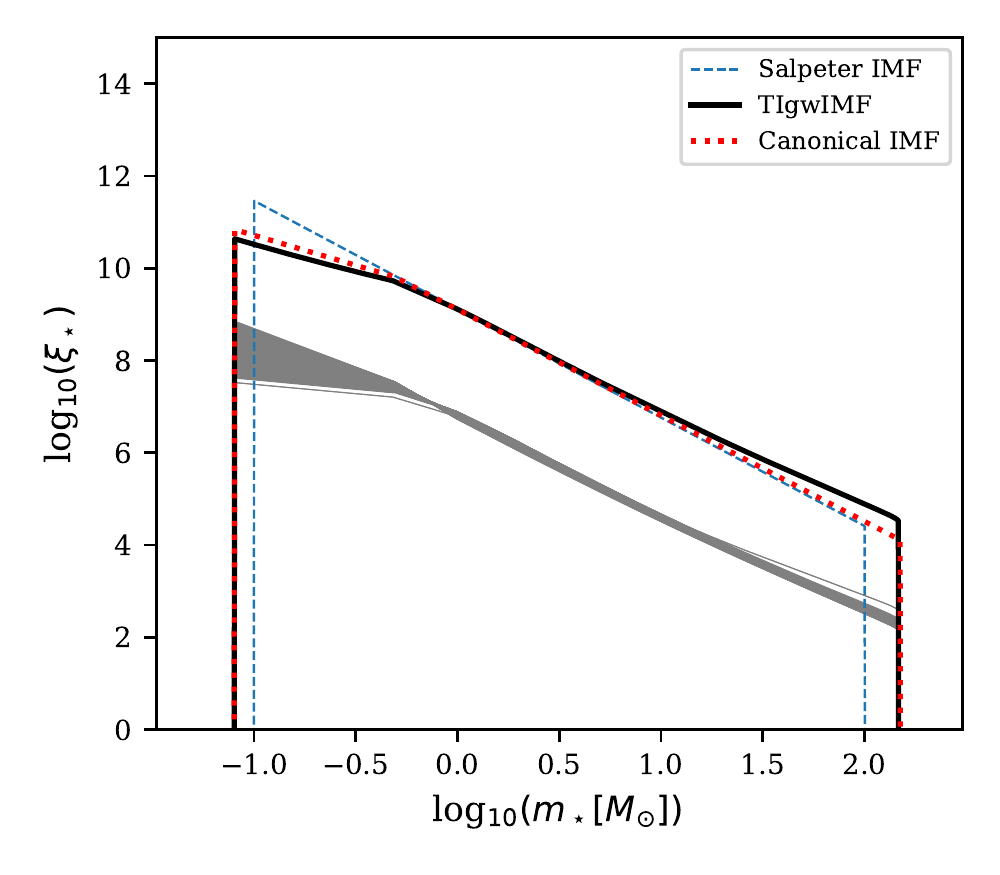}
    \caption{Same as Fig.~\ref{fig:galaxy_evolution_fig_TIgwIMF_82} but for a galaxy with a final mass of $M_{\rm dyn}\approx 10^{9.5} M_\odot$. This TIgwIMF is well approximated by the blue model in Fig.~\ref{fig:gwIMF_for_diff_SFR}.}
    \label{fig:galaxy_evolution_fig_TIgwIMF_217}
\end{figure}

Fig.~\ref{fig:galaxy_evolution_fig_TIgwIMF_127} shows the gwIMF for each 10 Myr star formation epoch and the TIgwIMF for a galaxy with $\mathrm{log}_{10}(\bar{\psi}_{\delta t}[M_\odot/\mathrm{yr}])=-0.2693$, $\tau_{\rm SF}=1.11~\mathrm{Gyr}$, $g_{\rm conv}=0.3204$ and a final mass of $10^{8.502}~M_\odot$, corresponding to the lowest-SFR green gwIMF in Fig.~\ref{fig:gwIMF_for_diff_SFR}.
Fig.~\ref{fig:galaxy_evolution_fig_TIgwIMF_217} shows the gwIMF for each 10 Myr star formation epoch and the TIgwIMF for a galaxy with $\mathrm{log}_{10}(\bar{\psi}_{\delta t}[M_\odot/\mathrm{yr}])=0.527$, $\tau_{\rm SF}=1.88~\mathrm{Gyr}$, $g_{\rm conv}=0.2796$ and a final mass of $10^{9.4837}~M_\odot$, corresponding to the blue gwIMF in Fig.~\ref{fig:gwIMF_for_diff_SFR}.

\end{appendix}

\end{document}